\documentclass{aa}  

\usepackage{graphicx}
\usepackage{txfonts}
\usepackage{lipsum}
\usepackage{subcaption}
\usepackage{lscape}
\usepackage{placeins}

\usepackage[colorlinks=true, linkcolor=blue, urlcolor=blue, citecolor=blue, anchorcolor=blue]{hyperref}

\usepackage{orcidlink}

\usepackage{subfiles}

\begin{document}

\title{Native-resolution retrievals of VHS 1256-1257 b spanning the JWST/NIRSpec wavelength range}
\subtitle{Chemical composition of a partially cloudy atmosphere}

\author{S. de Regt\inst{\ref{instLeiden}}\orcidlink{0000-0003-4760-6168}\corrauth{regt@strw.leidenuniv.nl}
    \and N. Whiteford\inst{\ref{instAMNH}}\orcidlink{0000-0001-8818-1544}\email{niallwhiteford@gmail.com}
    \and B. E. Miles\inst{\ref{instArizona}}\orcidlink{0000-0002-5500-4602}\email{bemiles@arizona.edu}
    \and S. Gandhi\inst{\ref{instWarwick},\ref{instCEH}}\orcidlink{0000-0001-9552-3709}\email{Siddharth.Gandhi@warwick.ac.uk}
    \and D. Gonz\'alez Picos\inst{\ref{instLeiden}}\orcidlink{0000-0001-9282-9462}\email{picos@strw.leidenuniv.nl}
    \and I. A. G. Snellen\inst{\ref{instLeiden}}\orcidlink{0000-0003-1624-3667}\email{snellen@strw.leidenuniv.nl}
    }
\institute{Leiden Observatory, Leiden University, P.O. Box 9513, 2300 RA, Leiden, The Netherlands\label{instLeiden}
    \and Department of Astrophysics, American Museum of Natural History, Central Park West at 79th Street, NY 10024, USA\label{instAMNH}
    \and Steward Observatory, University of Arizona, Tucson, AZ 85721, USA\label{instArizona}
    \and Department of Physics, University of Warwick, Coventry CV4 7AL, UK\label{instWarwick}
    \and Centre for Exoplanets and Habitability, University of Warwick, Gibbet Hill Road, Coventry CV4 7AL, UK\label{instCEH}
    }

\date{Received date / Accepted date}

\abstract
{The wide wavelength coverage and sensitivity offered by JWST enable detailed analyses of extrasolar atmospheres. At its highest resolution ($\mathcal{R}\sim$\,$2700$), the Near Infrared Spectrograph (NIRSpec) measures the absorption from atomic, molecular, and isotopic gases whose abundances reflect the chemical composition of accreted material, making them key tracers of the formation environment. Orbiting an inner M-dwarf binary at a wide separation, the planetary-mass companion VHS 1256-1257 b is one of the most variable sub-stellar objects known, with flux variations of $10$--$30\%$. }
{We aim to constrain the chemical composition of this companion in order to decipher its formation history and explore the origin of its extreme variability.}
{We analyse $0.97$--$5.27\ \mathrm{\mu m}$ (G140H, G235H, and G395H) JWST/NIRSpec spectra of VHS 1256-1257 b, from Early Release Science program \#1386. We update the data reduction and employ \texttt{petitRADTRANS} to carry out atmospheric retrievals at the native spectral resolution. We model a partially cloudy atmosphere in chemical disequilibrium and fit directly for the elemental and isotopic abundances.}
{Our best-fitting model closely matches the observations, bringing the residuals down to $\sim$\,$1\%$. The results are cautiously interpreted as degeneracies can bias some parameter constraints, most notably the mass and radius. Still, the retrieval finds a partial cloud deck covering $\sim$\,$79\%$ of the visible surface, with a clearer column dominating at short wavelengths. Small coverage changes of $\pm1$--$3\%$ can account for the high observed variability. From the many detected atmospheric gases, we infer a metallicity, $\mathrm{C/O}$ ratio, and $\mathrm{^{12}C/^{13}C}$ isotope ratio in line with a solar composition, while the $^{18}$O isotope appears depleted relative to the Sun and local interstellar medium. The minor isotope abundances are significantly lower than previous studies suggested, underlining the importance of our updated spectral extraction.}
{The $^{18}$O-depletion defies our understanding of VHS 1256-1257 b's assumed top-down formation. Our interpretation of the retrieved composition is complicated further by the lack of abundance measurements for the host stars. Nevertheless, this study demonstrates the value of panchromatic, native-resolution retrievals for characterising complex extrasolar atmospheres.}


\maketitle
\nolinenumbers

\section{Introduction}
The chemical composition of exoplanet atmospheres offers an insight into the material accreted during the formation and subsequent evolution. The chemical make-up of these building blocks is shaped by fractionation processes that occur throughout the circumstellar disk, thus encoding the birth environment into the exoplanet composition \citep{Oberg_ea_2021,Molliere_ea_2022}. Elemental abundance ratios such as the C/O, N/O, and refractory-to-volatile ratios have been proposed as valuable tracers of formation location \citep{Oberg_ea_2011,Madhusudhan_ea_2012,Turrini_ea_2021,Lothringer_ea_2021}, but isotope abundances from $^{13}$C, $^{18}$O, $^{15}$N, and D have been put forward as well \citep{Molliere_ea_2019a,Morley_ea_2019,Zhang_ea_2021a,Nomura_ea_2023}. The deuterium abundance is particularly useful since its depletion reveals that a sub-stellar object is sufficiently massive to have ignited deuterium-fusion, a criterion that is commonly -- but not universally -- used to distinguish giant planets from brown dwarfs \citep{Saumon_ea_2008,Spiegel_ea_2011,Morley_ea_2024}. Nevertheless, the low primordial deuterium-to-hydrogen ratio poses a major challenge to such measurements (e.g. protosolar $\mathrm{D/H}=(1.67^{+0.25}_{-0.25})\cdot10^{-5}$; \citealt{Asplund_ea_2021}).

The arrival of the James Webb Space Telescope (JWST; \citealt{Rigby_ea_2023}) has opened a new avenue for the detailed characterisation of extrasolar atmospheres. The high sensitivity and broad wavelength coverage offered by its Near Infrared Spectrograph (NIRSpec; \citealt{Boker_ea_2022}) and Mid InfraRed Instrument (MIRI; \citealt{Wells_ea_2015,Wright_ea_2023}) enable detailed analyses of atmospheric chemistry, thermal structure, and cloud properties on self-luminous exoplanets and brown dwarfs (e.g. \citealt{Biller_ea_2024,Hoch_ea_2025, McCarthy_ea_2025,Molliere_ea_2025,Zhang_ea_2025, Ruffio_ea_2026,Xuan_ea_2026}). At the highest resolutions, the spectrographs can distinguish the absorption lines of different isotopologues, resulting in detections of $^{13}$CO, C$^{18}$O, C$^{17}$O, H$_2^{18}$O, $^{15}$NH$_3$, and CH$_3$D (e.g. \citealt{Gandhi_ea_2023,Lew_ea_2024, Barrado_ea_2023,Rowland_ea_2024, Kuhnle_ea_2025,Gonzalez_Picos_ea_2025b}).

VHS J125601.92-125723.9 (hereafter VHS 1256-1257) is a hierarchical triple system with an inner M7.5+M7.5 binary on a tight orbit ($a=1.96\pm0.03\ \mathrm{AU}$; \citealt{Stone_ea_2016,Rich_ea_2016,Dupuy_ea_2023}) and an outer companion at a wide projected separation of $\sim$\,$8.06 \ \mathrm{arcsec}$ ($\sim$\,$380\ \mathrm{AU}$ at $d=21.15\pm0.21\ \mathrm{pc}$; \citealt{Gauza_ea_2015,Gaia_ea_2021}). The system age is estimated at $140\pm20\ \mathrm{Myr}$ which makes the luminosity of VHS 1256-1257 b consistent with both deuterium-inert and deuterium-fusing cooling tracks \citep{Dupuy_ea_2023,Miles_ea_2023}. The resulting mass distribution is bimodal with peaks near $12.0\pm0.1$ and $16\pm1\ M_\mathrm{Jup}$ \citep{Dupuy_ea_2023}, on either side of the $\sim$\,$13\ M_\mathrm{Jup}$ deuterium-burning limit. Hence, VHS 1256-1257 b can be categorised as both an exoplanet and a brown-dwarf companion, but here we opt instead for an intermediate term: Planetary-Mass Companion (PMC). Abundance measurements of deuterated isotopologues like HDO and CH$_3$D would help to better inform the mass of VHS 1256-1257 b \citep{Morley_ea_2019,Molliere_ea_2019a}. The VHS 1256-1257 system has an intriguing architecture, where the binary orbit, companion orbit and companion spin axis are all misaligned relative to each other \citep{Poon_ea_2024}. These misalignments, combined with the high mass and wide orbit of VHS 1256-1257 b, are unlikely to result from bottom-up planet formation \citep{Holzknecht_ea_2026}. Instead, \citet{Poon_ea_2024} propose a top-down formation scenario where core or filament fragmentation forms a misaligned multiple system that subsequently decays to tighten the separation of the inner VHS 1256-1257 AB binary. Constraining the atmospheric compositions within this system will help to shed light on the elemental and isotopic enrichment experienced during this formation pathway.

Since its discovery, various spectro-photometric observations have facilitated in-depth studies of the VHS 1256-1257 b atmosphere, yielding an effective temperature in the range of $1100$--$1400\ \mathrm{K}$, a low surface gravity ($3.25\lesssim\log_{10}\textit{g}\lesssim4.50$), but also a wide range of constraints for the metallicity ($-0.4\lesssim\mathrm{[M/H]}\lesssim+0.4$) and C/O ratio ($0.38\lesssim\mathrm{C/O}\lesssim0.63$; \citealt{Hoch_ea_2022,Dupuy_ea_2023,Miles_ea_2023, Petrus_ea_2023,Petrus_ea_2024,Radcliffe_ea_2026}). The relative absorption depths of CO and CH$_4$, respectively at $4.4$--$5.0$ and $3.1$--$3.5\ \mathrm{\mu m}$, indicate a chemical disequilibrium caused by the strong vertical mixing of gases \citep{Miles_ea_2018,Miles_ea_2023}. The reddened colours of VHS 1256-1257 b suggest a cloudy atmosphere \citep{Gauza_ea_2015}, which is corroborated by the strong $\sim$\,$10\ \mathrm{\mu m}$ silicate-cloud absorption detected in JWST/MIRI spectra (\citealt{Miles_ea_2023}, Whiteford et al., \textit{subm.}). Monitoring observations also reveal a strikingly high variability with peak-to-valley amplitudes of $>20\%$ (and up to $38\%$ over a $2\ \mathrm{yr}$ baseline; \citealt{Zhou_ea_2022}), making VHS 1256-1257 b one of the most variable sub-stellar objects known to date \citep{Bowler_ea_2020,Zhou_ea_2020, Zhou_ea_2022}. The observed variability likely results from the (dis)-appearance of non-uniform temperature and cloud structures as the PMC rotates \citep{Tan_ea_2025}. \citet{Zhou_ea_2020} find that the estimated rotation period of $21$--$24\ \mathrm{hr}$, together with a projected rotational velocity of $\textit{v}\sin i=13.5\pm4.1\ \mathrm{km\ s^{-1}}$ \citep{Bryan_ea_2018}, point towards an equator-on viewing geometry for VHS 1256-1257 b. Using additional high-resolution spectra, \citet{Poon_ea_2024} similarly infer a perpendicular spin-axis orientation at $i_\mathrm{p}=90\pm18^\circ$.

In this paper, we study the atmospheric chemistry and structure of VHS 1256-1257 b using high-fidelity spectra taken with JWST/NIRSpec as part of the Early Release Science Program \#1386 (ERS; \citealt{Hinkley_ea_2022}). Section~\ref{sect:methods} describes the data reduction and the modelling framework that we used to infer the atmospheric properties of this PMC. We present the results of the atmospheric retrieval in Sect.~\ref{sect:results}, and these are subsequently discussed in Sect.~\ref{sect:discussion} in the context of potential biases and the system's probable formation history. In Sect.~\ref{sect:conclusions}, we summarise the main conclusions drawn from this atmospheric investigation. 

\section{Methods} \label{sect:methods}
The primary goal of this work is to characterise the chemical composition of VHS 1256-1257 b's atmosphere. To that end, it is crucial to resolve molecular and atomic absorption lines and measure their depths. Near-infrared spectroscopy is particularly suitable for constraining the abundances of volatile molecules (e.g. CO, H$_2$O, CH$_4$; \citealt{Greene_ea_2016}), and we therefore focus our analysis on the JWST/NIRSpec spectra initially presented by \citet{Miles_ea_2023}. We refer the reader to Whiteford et al. (\textit{subm.}) for an atmospheric retrieval study of the combined NIRSpec and MIRI data from \citet{Miles_ea_2023}. Owing to the considerable computational demand, Whiteford et al. (\textit{subm.}) bin the spectra down to a resolving power of $300$. Although this makes the inference less sensitive to gaseous absorption, the broader features resulting from (silicate) clouds and temperature structures are characterised well. As such, the analysis of Whiteford et al. (\textit{subm.}) is very complementary to our focus on the atmospheric gases from the NIRSpec spectra at their native resolution.

Moreover, this study is a natural extension to \citet{Gandhi_ea_2023}, who conduct atmospheric retrievals to report the first JWST measurements of CO isotopologues in an extrasolar atmosphere. Where \citet{Gandhi_ea_2023} limit their analysis to the G395H-NRS2 grating-detector pair (i.e. $\sim$\,$4.1$--$5.3\ \mathrm{\mu m}$), here we consider the complete $0.97$--$5.27\ \mathrm{\mu m}$ wavelength range accessed by NIRSpec. This enables a comprehensive exploration of molecular and atomic absorbers, and probes the chemical disequilibrium and enrichment of VHS 1256-1257 b's atmosphere. In addition, the abundance constraints of \citet{Gandhi_ea_2023} provide a useful reference to assess the impact of our new data reduction.

\subsection{Data reduction} \label{sect:data_reduction}
We investigate the JWST/NIRSpec integral field spectra of VHS 1256-1257 b, acquired on July 5th, 2022 as part of ERS program \#1386 (PI: Hinkley; \citealt{Hinkley_ea_2022,Miles_ea_2023}). The three gratings, G140H/F100LP, G235H/F170LP, and G395H/F290LP, cover wavelengths between $0.97$--$5.27\ \mathrm{\mu m}$ at spectral resolutions of $\sim$\,$2000$--$4100$ \citep{Shajib_ea_2025}. The observations consist of four dither positions with a total exposure time of $1225.468\ \mathrm{s}$ per grating. \citet{Miles_ea_2023} employed the most up-to-date JWST pipeline (1.7.2) and CRDS context (977) for the initial data reduction, but numerous developments to the calibration data and software have greatly improved our reduction capabilities since then. We therefore repeated the data reduction using pipeline version 1.20.2 and CRDS context 1464, the latest available at the time of this analysis. 

\begin{figure*}[ht!]
    \centering
    \includegraphics[width=17cm]{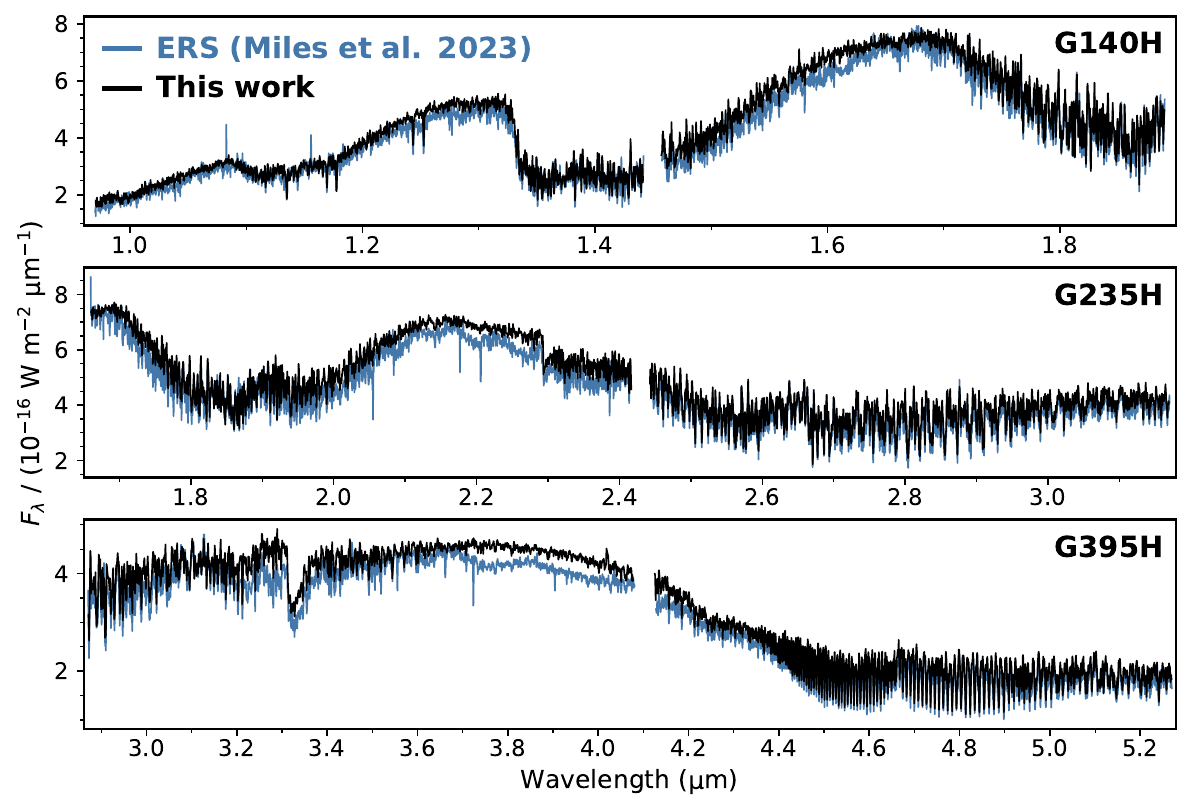}
    \caption{Comparison between the NIRSpec data of VHS 1256-1257 b presented by \citet{Miles_ea_2023} and the data reduction carried out in this work. The new spectrum contains fewer outliers and minimises the oscillations seen in regions of the continuum.}
    \label{fig:ERS_comparison}
\end{figure*}

Starting from the uncalibrated data on MAST, the Stage 1 pipeline performs detector-level corrections, including ${\rm1/f}$ noise removal, and converts detector ramps to slopes. The Stage 2 step then applies instrumental corrections, calibrates the flux and produces 3D spectral cubes for each of the four dither positions. We built the spectral cubes in the ``ifualign'' coordinate system in order to subtract residual background signal, similar to the MIRI/MRS correction carried out by \citet{Matthews_ea_2025}. Analogous to \citet{Miles_ea_2023}, we used a custom Python routine\footnote{\url{https://github.com/samderegt/JWST_VHS1256b_reduction_v3}} to extract 1D spectra for each grating, detector, and dither. First, the spectral cube is median-combined along the wavelength axis and the four Point-Spread Functions (PSFs) are fitted using 2D Gaussians, allowing us to determine the size and location of the circular extraction apertures. A radius of $4\sigma$ optimised the enclosed flux while also avoiding many noisy pixels and outliers. Using the \texttt{photutils} package \citep{Bradley_ea_2025}, we extracted the flux at each wavelength channel and thus constructed a 1D spectrum for each dither. Pixels that significantly deviated from other dithers were masked and the four 1D spectra were subsequently combined through a weighted average. As a final step, we performed an aperture correction to account for the $\sim$\,$7$--$10\%$ of flux lost outside of the $4\sigma$ extraction radius. We estimated a linear, wavelength-dependent correction factor by comparing the spectrum with an extraction that utilised a larger $10\sigma$ radius.

Figure~\ref{fig:ERS_comparison} compares the reduced spectrum from this work to the initial ERS reduction of \citet{Miles_ea_2023}. The new spectrum contains fewer outliers and no longer shows the beating pattern that was visible in the continuum near $\sim$\,$2.2$ and $3.8\ \mathrm{\mu m}$. Furthermore, the aperture correction yields an overall higher flux in all NIRSpec gratings and detectors. The median absolute deviations vary from $\sim$\,$5$ (G140H-NRS1) to $25$ (G395H-NRS1) times the flux uncertainty, but can be as high as $\sim$\,$50\sigma$ near $3.2$--$3.4\ \mathrm{\mu m}$. The median signal-to-noises of the new spectrum range from $\sim$\,$90$ (G140H-NRS1) to $380$ (G395H-NRS1) per pixel.

\subsection{Atmospheric retrieval framework}
To infer the atmospheric properties of VHS 1256-1257 b, we employed a Bayesian retrieval framework. The \texttt{petitRADTRANS} radiative transfer code (version 3.1; \citealt{Molliere_ea_2019,Molliere_ea_2020,Blain_ea_2024}) is used to generate emission spectra from a set of variables and parametrisations, which are sampled with the \texttt{PyMultiNest} nested sampling algorithm \citep{Feroz_ea_2009,Buchner_ea_2014}. We used 1000 live points at a constant sampling efficiency of $5\%$ to ensure a consistent convergence of the parameters.

We defined the likelihood as
\begin{align}
    \ln\mathcal{L} &= -\frac{1}{2}\sum_i \bigg(\ln(2\pi\sigma_i^2) + \Big(\frac{d_i-m_i}{\sigma_i}\Big)^2 \bigg),
\end{align}
where $d_i$ and $m_i$ are the data and model fluxes, respectively, and the sum is performed over all pixels, detectors and gratings. The flux errors $\sigma_i^2=\sigma_{\mathrm{obs},i}^2+10^b$ include both the observational uncertainties, resulting from the data reduction, as well as an inflation factor that compensates for potential error-underestimation or model deficiencies which could otherwise lead to biased parameter constraints \citep{Line_ea_2015}. We chose to adopt grating- and detector-dependent inflations to better account for the varying fit quality at different wavelengths. In the retrievals, we sample three independent $10^{b_\mathrm{grat}}$ terms with uniform priors $\mathcal{U}(0.01,100)$. The flux errors in each grating-detector pair are then inflated using the median of the variance vector via
\begin{align}
    \sigma_i^2 &= \sigma_{\mathrm{obs},i}^2+10^{b_\mathrm{grat}}\cdot\mathrm{median}(\vec{\sigma}_\text{grat-det}^2). \label{eq:uncertainty}
\end{align}

\subsubsection{General properties}
Based on the bolometric luminosity and estimated age of VHS 1256-1257 b, cooling-track models find masses consistent with a giant exoplanet ($\sim$\,$12\ M_\mathrm{Jup}$) but also a deuterium-burning brown dwarf ($\sim$\,$16\ M_\mathrm{Jup}$; \citealt{Miles_ea_2023, Dupuy_ea_2023}). To cover both scenarios, we employ a uniform prior between $10$ and $20\ M_\mathrm{Jup}$. The mass is combined with the radius, which is retrieved between $1$ and $1.7\ R_\mathrm{Jup}$ (based on previous estimates of $\sim$\,$1.3\ R_\mathrm{Jup}$; \citealt{Dupuy_ea_2023,Miles_ea_2023,Petrus_ea_2024}; Whiteford et al., \textit{subm.}), to compute the surface gravity through $\textit{g}=GM/R^2$. The radius is also used to scale the emitted flux by $(R/d)^2$, where the distance is $d=21.15\pm0.21\ \mathrm{pc}$ \citep{Gaia_ea_2021}. Furthermore, we fit for the radial velocity of VHS 1256-1257 b relative to the barycentric rest frame, following the barycentric correction applied by the JWST pipeline. We also use a constant projected rotational velocity of $v\sin{i}=8.7\ \mathrm{km\ s^{-1}}$, which is the median value measured from high-resolution spectra ($\sigma_{v\sin{i}}=0.1\ \mathrm{km\ s^{-1}}$; \citealt{Poon_ea_2024}). We use the \texttt{broadpy} package\footnote{\url{https://github.com/DGonzalezPicos/broadpy}} to convolve our model spectra to the instrumental resolutions of the three NIRSpec gratings. The wavelength-dependent resolving powers are adopted from \citet{Shajib_ea_2025}, who use in-flight data to fit polynomials that are $1$--$24\%$ higher than the pre-launch estimates.

\subsubsection{Temperature profile}
Following the parametrisation introduced by \citet{Zhang_ea_2023}, we model the thermal profile by retrieving the temperature gradient $\nabla_i=(d\ln{T_i})/(d\ln{P_i})$ at seven points in pressure, namely $100,10,...,10^{-4}\ \mathrm{bar}$. The \texttt{pRT} model is divided into 31 atmospheric layers with uneven spacing between $10^{2}$ and $10^{-4}\ \mathrm{bar}$, in order to have sufficient resolution near the photosphere ($\sim$\,$10$--$10^{-2}\ \mathrm{bar}$) while keeping computation times to a minimum. The temperature gradients $\nabla_j$ are obtained through linear interpolation between the knot-values $\nabla_i$. The temperature at each layer can then be computed as
\begin{align}
    T_j &= T_{j-1}\cdot\left(\frac{P_j}{P_{j-1}}\right)^{\nabla_j}.
\end{align}
This definition requires the retrieval of an anchor point for the temperature, $T_1$, which we determine at $P_1=10\ \mathrm{bar}$. We incorporate the expected radiative-convective equilibrium by choosing informed, Gaussian priors of \citet{Zhang_ea_2025} for the temperature gradients (see Table~\ref{tab:parameters}). The gradient priors of the upper atmosphere are centred near $\nabla_i\sim$\,$0$, and could therefore facilitate possible thermal inversions. 

\subsubsection{Clouds}
Similar to the approach of Whiteford et al. (\textit{subm.}), \citet{Radcliffe_ea_2026}, and other retrieval studies (e.g. \citealt{Vos_ea_2023,Molliere_ea_2025,Nasedkin_ea_2025,Zhang_ea_2025}), we model the emitted flux as the sum of two atmospheric columns with different cloud structures. To simulate different coverage fractions ($\mathcal{CF}$) for the two regions, we weigh them as
\begin{align}
    F_\mathrm{tot} &= \mathcal{CF}\cdot F_\mathrm{1} + (1-\mathcal{CF})\cdot F_\mathrm{2}.
\end{align}
The high photometric variability of VHS 1256-1257 b suggests that the cloud coverage may change for consecutive observations as a result of its rotation \citep{Bowler_ea_2020,Zhou_ea_2020,Zhou_ea_2022}. The three NIRSpec gratings were observed sequentially in the order G235H, G395H, G140H, over a time period of $1.5$ hours. We test for variability by defining two deviations ($\Delta\mathcal{CF}_\mathrm{G140H}$, $\Delta\mathcal{CF}_\mathrm{G235H}$) from the coverage in the central observation (i.e. G395H). Both columns are able to become dominant, but we limit the prior volume and avoid unphysically large fluctuations by imposing a Gaussian prior of $\mathcal{N}(\mu=0,\sigma=0.1)$ on the deviations $\Delta\mathcal{CF}_\mathrm{G140H}$ and $\Delta\mathcal{CF}_\mathrm{G235H}$.

In contrast to the silicate absorption in the MIRI spectra, the NIRSpec wavelength range does not cover distinctive cloud features. This means that our NIRSpec-only analysis cannot constrain the chemical composition of the condensates, while imposing specific cloud properties risks biasing the retrieved solution instead. We therefore choose to model the clouds with simplified non-gray opacities, which has the added benefit of reducing the number of free parameters compared to a more physically-motivated model (e.g. \citealt{Molliere_ea_2025}). The total opacity of one cloud deck is described with
\begin{align}
    \kappa_\mathrm{tot}(P,\lambda) &= \begin{cases}
        \kappa_\mathrm{base}\left(\dfrac{P}{P_\mathrm{base}}\right)^{f_\mathrm{sed}}\left(\dfrac{\lambda}{1\ \mathrm{\mu m}}\right)^{\xi} & P\leq P_\mathrm{base}, \\
        0 & P > P_\mathrm{base}, 
        \end{cases}
\end{align}
where $\kappa_\mathrm{base}$ and $P_\mathrm{base}$ are the opacity and pressure at the cloud base. The decay above this base is controlled with $f_\mathrm{sed}$ and $\xi$ modulates the wavelength-dependence. Following \citet{Molliere_ea_2020}, the absorption and scattering opacities are calculated using a single-scattering albedo, $\omega$, using
\begin{align}
    \kappa_\mathrm{abs}(P,\lambda) &= (1-\omega)\cdot\kappa_\mathrm{tot}(P,\lambda),  \\
    \kappa_\mathrm{scat}(P,\lambda) &= \omega\cdot\kappa_\mathrm{tot}(P,\lambda). 
\end{align}
The two atmospheric columns share a cloud deck and an additional cloud is fit in column 2. Both clouds are retrieved with the same priors (see Table~\ref{tab:parameters}). As such, this model can mimic a deep iron cloud with a patchy silicate cover atop, as has been found in previous retrieval analyses of sub-stellar objects \citep{Vos_ea_2023,Zhang_ea_2025,Molliere_ea_2025} and VHS 1256-1257 b (Whiteford et al., \textit{subm.}). 

\subsubsection{Chemistry}
We employ a chemical disequilibrium model to obtain volume-mixing ratios of the molecular and atomic gases present in the atmosphere. As described in \citet{de_Regt_ea_2026}, this approach uses a reduced input to the \texttt{FastChem} code \citep{Stock_ea_2018,Stock_ea_2022,Kitzmann_ea_2024,Kitzmann_ea_2026}, thereby enabling computations of equilibrium abundances for each model during the retrieval. In addition, this method accounts for rainout condensation and directly constrains the elemental enrichments, $[{\rm X/H}]$, relative to the solar composition \citep{Asplund_ea_2021}. Here, we fit for the elemental abundances of C, O, N, S, F, K+Na, and Cr+Fe, because we detect them as constituents of molecular or atomic gases (see Sect.~\ref{sect:cross_correlation}). We choose to jointly retrieve the enhancements of K and Na, as well as Cr and Fe, to reduce the number of free parameters with elements that have a limited spectral impact only across narrow wavelength ranges. Other metals are assumed to have solar abundances since we lack direct measurements and aim to avoid biasing the abundances of the detected gases. It should be noted, however, that the availability of Mg and Si affects the degree of oxygen trapping into silicate clouds \citep{Calamari_ea_2024}, and thus the abundances of oxygen-bearing molecules such as CO and H$_2$O. We leave it to future studies to explore whether these refractory elements can be inferred from gaseous molecular abundances alone.

The volume-mixing ratios of H$_2$O, CH$_4$, CO, NH$_3$, and HCN are held constant at $P<P_\mathrm{quench}$ to simulate a mixing-induced disequilibrium. These quench points ($P_\mathrm{quench,\ H_2O-CH_4-CO}$, $P_\mathrm{quench,\ NH_3}$, $P_\mathrm{quench,\ HCN}$) are located where the chemical timescale of the limiting reaction (adopted from \citealt{Zahnle_ea_2014}) equates to the vertical-mixing timescale. The CO$_2$ abundance also gets quenched, but will first be enhanced due to the deeper quenching of CO. We therefore adopt the CO$_2$ abundance prescription outlined by \citet{Wogan_ea_2025}. The mixing timescale is defined as
\begin{align}
    t_{\rm mix}(P) &= \frac{\left(H(P)\right)^2}{K_{\rm zz}(P)} = \frac{1}{K_{\rm zz}(P)}\left(\frac{k_B\cdot T(P)}{\mu(P)\cdot m_{\rm p}g}\right)^2, \label{eq:t_mix}
\end{align}
where the scale height $H$ is used as the mixing length and depends on the temperature $T$, mean molecular weight $\mu$, and surface gravity $\textit{g}$, which vary with each evaluated model. The eddy diffusion coefficient $K_{\rm zz}$ regulates the mixing efficiency and is retrieved via three free parameters in a step profile as
\begin{align}
    K_\mathrm{zz}(P) &= \begin{cases}
        K_\mathrm{zz,upper} & P \leq P_{K_\mathrm{zz}}, \\
        K_\mathrm{zz,lower} & P > P_{K_\mathrm{zz}},
        \end{cases}
\end{align}
where $P_{K_\mathrm{zz}}$ defines the transition pressure. This parameterisation can reproduce the orders-of-magnitude decrease that is expected when sub-stellar atmospheres transition from convection-dominated energy transport to a radiative regime (e.g. \citealt{Mukherjee_ea_2022,Mukherjee_ea_2024}). The shorter chemical timescales of CO$_2$ will intersect the mixing timescale at a lower quench pressure than CO, CH$_4$, H$_2$O, NH$_3$, and HCN \citep{Zahnle_ea_2014}, thus making CO$_2$ a suitable molecule to test altitude variations of $K_\mathrm{zz}$ \citep{Lew_ea_2026}. 

Finally, we fit for the isotope ratios of $\mathrm{^{12}C/^{13}C}$, $\mathrm{^{16}O/^{18}O}$, $\mathrm{^{16}O/^{17}O}$, and $\mathrm{H/D}$, which are used to calculate the volume-mixing ratios of $\mathrm{^{13}CH_4}$, $\mathrm{^{13}CO_2}$, $\mathrm{H_2^{18}O}$, $\mathrm{^{16}OC^{18}O}$, $\mathrm{H_2^{17}O}$, HDO, and $\mathrm{CH_3D}$ via
\begin{alignat}{2}
    &\mathrm{^{13}CH_4} = \left(\mathrm{\dfrac{^{12}C}{^{13}C}}\right)^{-1}\mathrm{CH_4}, \quad && \mathrm{^{13}CO_2} = \left(\mathrm{\dfrac{^{12}C}{^{13}C}}\right)^{-1}\mathrm{CO_2}, \\
    &\mathrm{H_2^{18}O} = \left(\mathrm{\dfrac{^{16}O}{^{18}O}}\right)^{-1} \mathrm{H_2O}, \quad && \mathrm{^{16}OC^{18}O} = 2\cdot \left(\mathrm{\dfrac{^{16}O}{^{18}O}}\right)^{-1} \mathrm{CO_2}, \\
    &\mathrm{H_2^{17}O} = \left(\mathrm{\dfrac{^{16}O}{^{17}O}}\right)^{-1} \mathrm{H_2O}, \\
    &\mathrm{HDO} = 2\cdot\left(\mathrm{\dfrac{H}{D}}\right)^{-1} \mathrm{H_2O}, \quad && \mathrm{CH_3D} = 4\cdot\left(\mathrm{\dfrac{H}{D}}\right)^{-1} \mathrm{CH_4}.
\end{alignat}
The multiplicative factors of $2$ and $4$ (or $1$) account for the number of atoms that can be replaced by the heavier isotope within each molecule. In addition to these isotope ratios, we retrieve separate isotopologue ratios for $\mathrm{^{12}CO/^{13}CO}$, $\mathrm{C^{16}O/C^{18}O}$, and $\mathrm{C^{16}O/C^{17}O}$ to enable a direct comparison with \citet{Gandhi_ea_2023}. The CO isotopologues are reliably detected, but our test retrievals revealed that the $^{13}$C abundance could be biased when jointly retrieved with the $^{13}$CH$_4$ and $^{13}$CO$_2$ isotopologues (see Sect.~\ref{sect:composition}). Apart from the minor isotopologues, we include the line opacities of CO, H$_2$O, CH$_4$, CO$_2$, NH$_3$, H$_2$S, HF, HCN, CrH, FeH, SiO, K, and Na, calculated with \texttt{pyROX}\footnote{\url{https://py-rox.readthedocs.io}} \citep{de_Regt_ea_2025b}. Table~\ref{tab:opacity_refs} lists the references for the utilised line-opacity data. Furthermore, we account for the collision-induced absorption of H$_2$-H$_2$ and H$_2$-He pairs \citep{Borysow_ea_1988,Borysow_ea_1989,Richard_ea_2012} as well as Rayleigh scattering from H$_2$ and He \citep{Dalgarno_ea_1962,Chan_ea_1965}. We reduce computation times by downsampling the line opacities by factors of $9$--$11$, depending on the grating-detector pair. Moreover, each grating-detector pair loads a pre-defined set of molecular and atomic opacities which are relevant at the covered wavelengths. Similarly, weakly-absorbing molecules and isotopologues are excluded at short wavelengths in column 2 as they are blanketed by the additional cloud deck.

\section{Results} \label{sect:results}
\begin{figure*}[ht!]
    \centering
    \includegraphics[width=17cm]{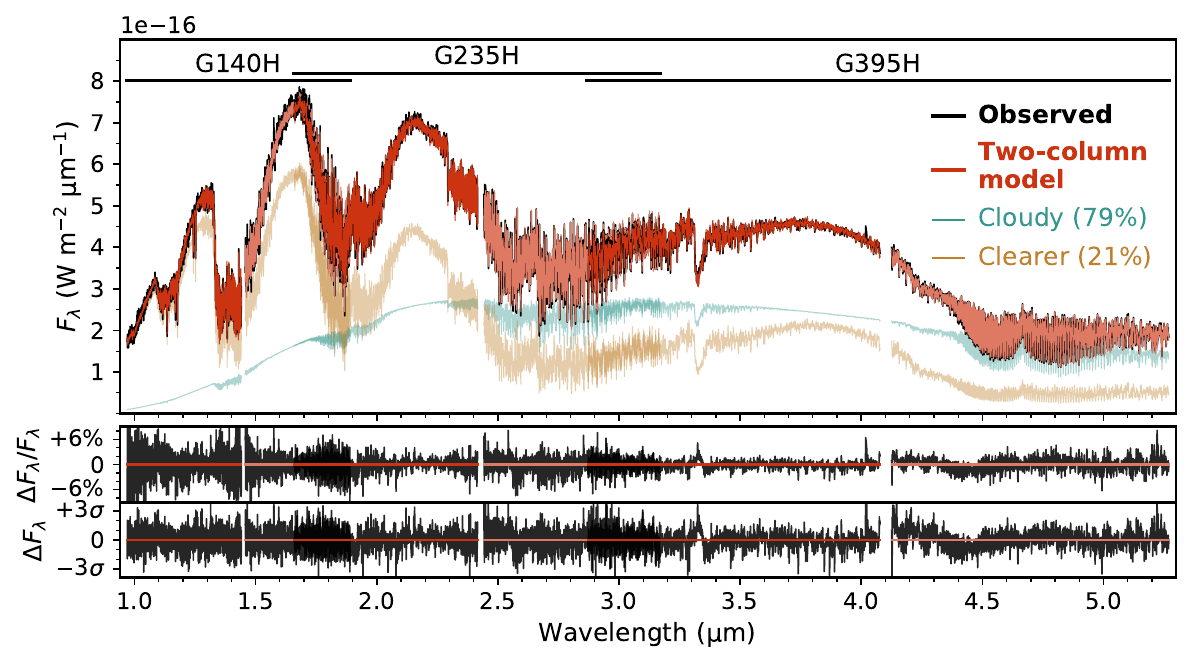}
    \caption{Best-fitting model to the NIRSpec data of VHS 1256-1257 b. The model spectrum (red) is comprised of two columns, a cloudy and clearer atmosphere, whose individual contributions are shown in cyan and gold, respectively. Despite covering $\sim$\,$21\%$ of the modeled surface, the clearer atmosphere dominates the spectrum at wavelengths $<2.5\ \mathrm{\mu m}$ and the observed absorption lines also primarily originate from this patch. The lower panels show the residuals relative to the total flux and error defined in Eq.~\ref{eq:uncertainty}.}
    \label{fig:bestfit}
\end{figure*}

The retrieval framework provides a good fit to the spectrum of VHS 1256-1257 b, as is visible in Fig.~\ref{fig:bestfit}. The zoomed-in panels of Fig.~\ref{fig:bestfit_zooms} show that the median absolute deviations of the residuals are typically at the $\sim$\,$1\%$ level and at some wavelengths as low as $\sim$\,$0.5\%$. The model spectrum reproduces almost all spectral line features, but small deviations can be seen near $2.55\text{--}2.70$, $3.30\text{--}3.35$ (CH$_4$ Q-branch), and $4.35\text{--}4.50\ \mathrm{\mu m}$. We speculate that these residuals are caused by (1) incomplete opacity due to deficient line-list data at high temperatures, or unaccounted absorbing species, or (2) a fundamental limitation of our parametrised model which consists of only two atmospheric columns that share thermo-chemical properties. 

Retrievals on the individual gratings can moderately improve the fits ($\lesssim$\,$0.1\%$ in MAD), but we find that the parameter constraints can be substantially biased depending on the spectral information considered. For example, the retrieved mass ($10$--$18\ M_\mathrm{Jup}$), radius ($1.0$--$1.5\ R_\mathrm{Jup}$), and by extension the derived surface gravity ($10^{4.2}$--$10^{4.4}\ \mathrm{cm\ s^{-2}}$), vary notably between the three individual gratings as well as the jointly fitted model, as is visible from the posterior distributions in Fig.~\ref{fig:corner_gratings}. For some parameters it is evident which gratings drive the constraints of the combined retrieval. For example, the CO isotopologue ratios are primarily inferred from the G395H grating because it spans the stronger fundamental absorption bands, contrary to G140H and G235H. On the other hand, these two bluer gratings probe deeper into the atmosphere and are more sensitive to clouds, resulting in narrower constraints on the cloud-coverage fraction. The joint retrieval benefits from the complementary sensitivities of all three gratings, which helps to alleviate the biases found in the separate retrievals. As such, we consider the combined model to be the most reliable and comprehensive characterisation of VHS 1256-1257 b's atmosphere, despite the small loss in fit quality.

\subsection{Chemistry}
\subsubsection{Detection of molecules and atoms} \label{sect:cross_correlation}
Through their unique absorption signatures, we find evidence for a large number of molecular and atomic gases. To assess the detection significances, we perform a cross-correlation using the residuals of the data ($d$) compared to a model without species X ($m_\mathrm{wo\ X}$) and a template with the modeled contribution of X (i.e. $d-m_\mathrm{wo\ X}$ vs $m-m_\mathrm{wo\ X}$, where $m$ is the model with all species; \citealt{Zhang_ea_2021b}). The cross-correlation coefficient at a velocity $v$ is then calculated via
\begin{align}
    \mathrm{CCF}(v) &= \sum_i \frac{\left(d_i-m_{\mathrm{wo\ X},i}\right) \left(m_i(v)-m_{\mathrm{wo\ X},i}(v)\right)}{\sigma_i^2}. \label{eq:CCF}
\end{align}
The cross-correlation function (CCF) is subsequently normalised to a signal-to-noise using the standard deviation of samples outside of the expected peak,
\begin{align}
    \mathrm{CCF}_\sigma(v) &= \frac{\mathrm{CCF}(v)}{\mathrm{std}\left((\mathrm{CCF}-\mathrm{ACF})\big|_{|v|>300\ \mathrm{km\ s^{-1}}}\right)}, \label{eq:CCF_SNR}
\end{align}
where we first subtract the template's auto-correlation to remove secondary aliasing peaks that may lead to under-estimated significances \citep{de_Regt_ea_2024}.

\begin{figure}[h!]
    \centering
    \includegraphics[width=\hsize]{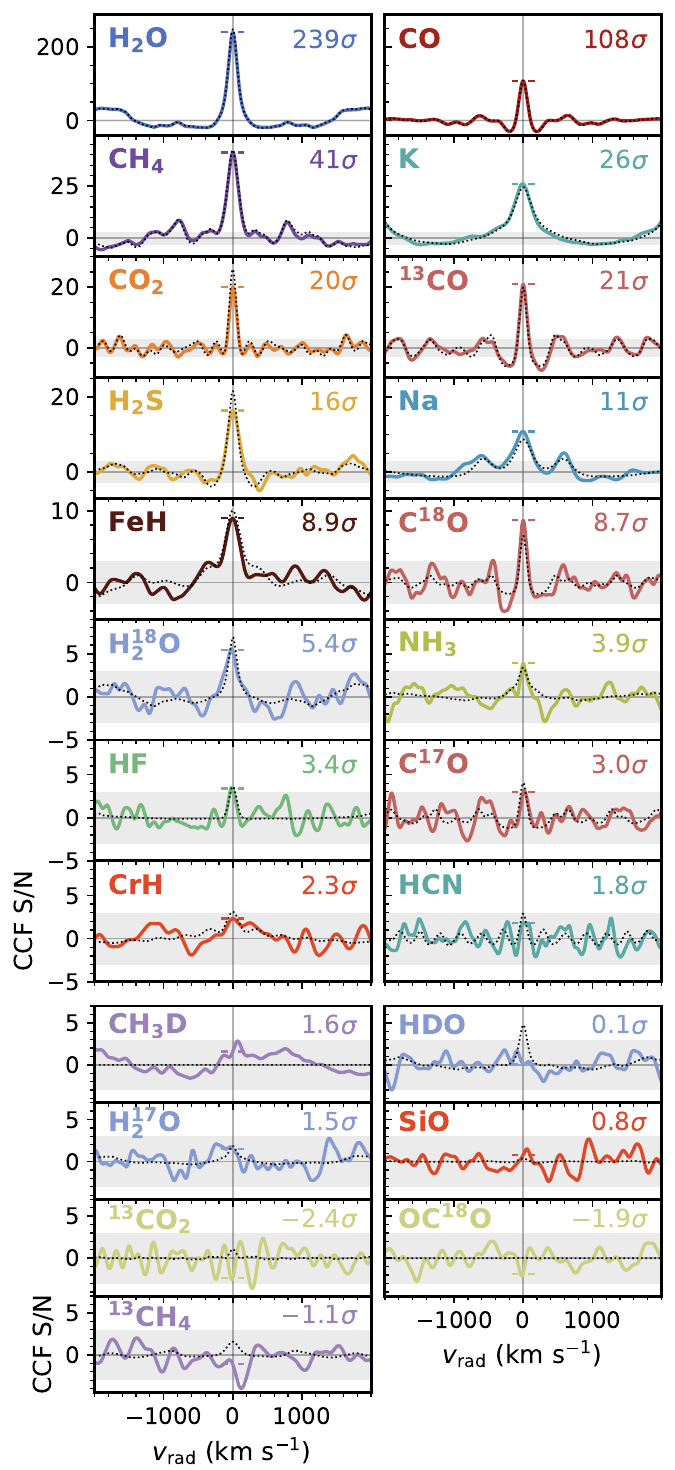}
    \caption{Cross-correlation functions of molecular and atomic absorbers, calculated with Eqs~\ref{eq:CCF} and~\ref{eq:CCF_SNR}. The dotted black lines show the auto-correlation of the model template with itself. The detection significances at $v=0\ \mathrm{km\ s^{-1}}$ are indicated in the upper right corners. Each panel-row uses a different y-axis limit for legibility.}
    \label{fig:CCF}
\end{figure}

Figure~\ref{fig:CCF} presents the cross-correlation functions for the 23 considered species, approximately ordered from strong detections to tentative indications and non-detections. In line with the visual identification of \citet{Miles_ea_2023}, we detect H$_2$O, CO, CH$_4$, CO$_2$, K, and Na through cross-correlation at $>3\sigma$. Their auto-correlation functions (black dotted lines) closely trace the height of the peaks, which implies that the cross-correlation signal is not notably inflated by unrelated aligned spectral residuals. We also reproduce the detections of $^{13}$CO and C$^{18}$O reported by \citet{Gandhi_ea_2023}, and find a marginal detection ($3.0\sigma$) for C$^{17}$O. NH$_3$ lines are also present in these observations, but they are more readily identified in the MIRI spectra of VHS 1256-1257 b (Whiteford et al., \textit{subm.}; \citealt{Malin_ea_2026}). For the first time in VHS 1256-1257 b, we detect the absorption of H$_2$S, H$_2^{18}$O, FeH, HF, and weak hints of CrH and HCN, but the lines of CH$_3$D, HDO, H$_2^{17}$O, SiO, $^{13}$CO$_2$, $^{16}$OC$^{18}$O, and $^{13}$CH$_4$ do not show significant cross-correlation peaks. 

\subsubsection{Elemental and isotopic composition} \label{sect:composition}
\begin{figure*}[ht!]
    \centering
    \includegraphics[width=16cm]{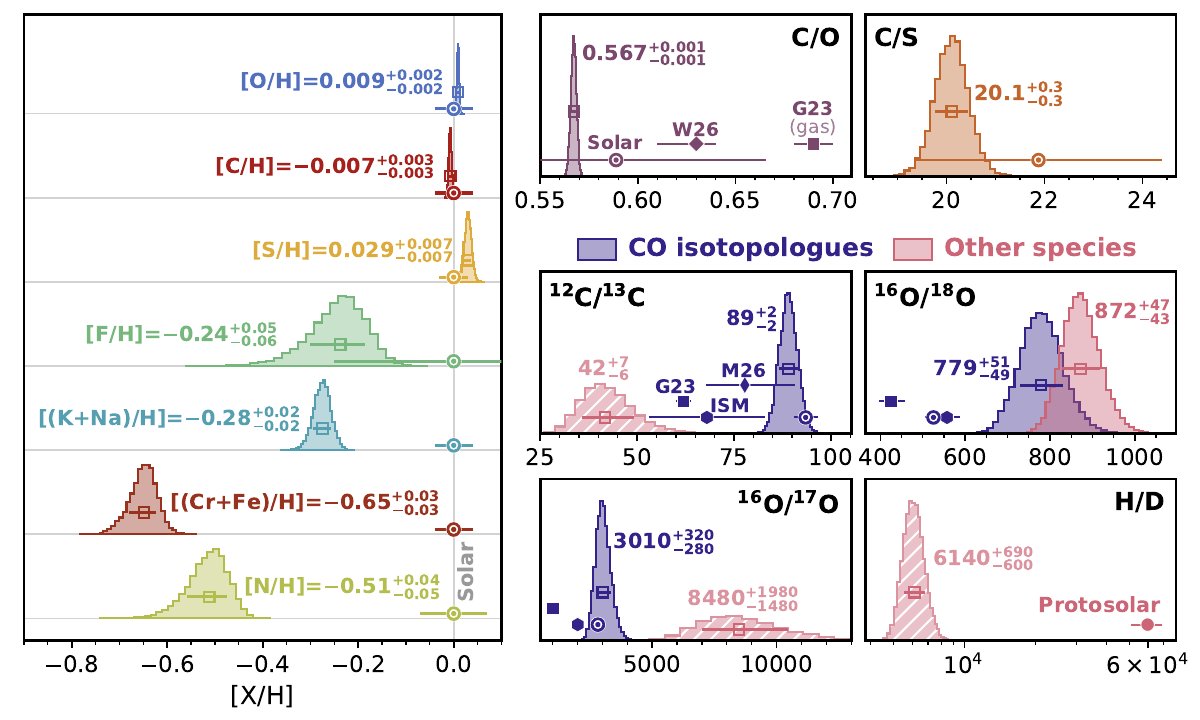}
    \caption{Elemental and isotopic composition retrieved for the atmosphere of VHS 1256-1257 b. \textit{Left panel}: Abundances relative to the solar composition compiled by \citet{Asplund_ea_2021}. \textit{Right panels}: Elemental and isotopic ratios of carbon, oxygen, sulphur, and deuterium. We show the isotopic constraint from the CO isotopologues in indigo, and those from other molecules in rose. The hatched posteriors specify the isotopes constrained without cross-correlation detections (see Sect.~\ref{sect:cross_correlation}). As references, we indicate the (proto)-solar ratios \citep{Lyons_ea_2018,Asplund_ea_2021}, local interstellar-medium ratios (ISM; \citealt{Milam_ea_2005,Wilson_1999}), as well as the VHS 1256-1257 b constraints from \citet{Gandhi_ea_2023}, Whiteford et al. (\textit{subm.}) and \citet{Malin_ea_2026} as G23, W26 and M26, respectively.}
    \label{fig:composition}
\end{figure*}

The depths of the molecular and atomic absorption lines allow us to infer the abundances of the constituent elements and isotopes. The compositional constraints for VHS 1256-1257 b are summarised in Fig.~\ref{fig:composition} and the posterior distributions of selected parameters are shown in the corner plot of Fig.~\ref{fig:corner}. To our knowledge, there are no elemental or isotopic abundance constraints for the inner M-dwarf binary, VHS 1256-1257AB. Unfortunately, this means that we cannot make a direct comparison of possible chemical enrichments (or depletions) relative to the host stars. Instead, we use the solar composition as a reference.

The left panel of Fig.~\ref{fig:composition} shows the retrieved elemental abundances with respect to the solar composition ($\mathrm{[X/H]}=0$). The posteriors of carbon, oxygen, and sulphur are tightly constrained due to the well-detected molecular absorption of CO, H$_2$O, CH$_4$, and H$_2$S (see Fig.~\ref{fig:CCF}). Accounting for the uncertainty reported by \citet{Asplund_ea_2021}, we find that the elemental abundances of $\mathrm{[C/H]}$, $\mathrm{[O/H]}$, $\mathrm{[S/H]}$, and $\mathrm{[F/H]}$ are consistent to within $\sim$\,$1\sigma$ of the solar values. While the $\mathrm{[N/H]}$, $\mathrm{[(K+Na)/H]}$, and $\mathrm{[(Cr+Fe)/H]}$ are systematically sub-solar, we caution that these constraints are likely biased toward low-metallicity solutions instead of reflecting true depletions, as discussed further in Sect.~\ref{sect:abundance_correlation}. The sub-solar bias for these elements is suspected to arise from their relatively low detection significances (see Fig.~\ref{fig:CCF}) and the limited wavelength coverage of the constraining atoms and molecules, in contrast to the more informative constraints provided by carbon, oxygen, and sulphur.

The right-hand panels of Fig.~\ref{fig:composition} show the elemental and isotopic abundance ratios inferred for VHS 1256-1257 b. Our retrieval constrains the carbon-to-oxygen ratio to $\mathrm{C/O}=0.567^{+0.001}_{-0.001}$, thus aligning with the solar value of $\mathrm{C/O}_\odot=0.59\pm0.08$ \citep{Asplund_ea_2021}. Similarly, the inferred carbon-to-sulphur ratio of $\mathrm{C/S}=20.1^{+0.3}_{-0.3}$ is consistent with the solar ratio ($\mathrm{C/S}_\odot=21.9\pm2.5$; \citealt{Asplund_ea_2021}). Recently, \citet{Xuan_ea_2026} reported a potential decreasing trend of $\mathrm{C/S}$ with orbital distance in the HR 8799 system, implying enhanced solid accretion further out. Our constraint for VHS 1256-1257 b is most compatible with the outer planet, HR 8799 b, measured at $\mathrm{C/S}_\mathrm{HR\ 8799\ b}=16.5\pm0.2$ \citep{Xuan_ea_2026}.

The four lower-right panels of Fig.~\ref{fig:composition} present isotope ratios inferred either from the isotopologues of CO (indigo), or other molecular species (rose) such as H$_2^{18}$O, H$_2^{17}$O, and HDO. We note that the $\mathrm{^{12}C/^{13}C}$, $\mathrm{^{16}O/^{17}O}$, and $\mathrm{H/D}$ constraints from the other molecules (hatched posteriors) should be cautiously interpreted in light of the cross-correlation non-detections found in Sect.~\ref{sect:cross_correlation}. Instead of measuring relative line depths for these isotopologues, the retrieval likely uses their weak opacity to adjust the spectral slope, which is more prone to bias from the specific data reduction and model parametrisation (see Sect.~\ref{sect:correlation}). On the other hand, the CO-isotopologue and $\mathrm{^{16}O/^{18}O}$ ratios (primarily from H$_2^{18}$O) are constrained by identifiable absorption lines, albeit marginally for C$^{17}$O. This could clarify the apparent disagreement between the inferred $^{13}$C and $^{17}$O ratios, whereas the two $^{18}$O ratios are $\sim$\,$2\sigma$ consistent with each other. The absence of line detections may also explain the elevated deuterium abundance of $\mathrm{H/D}=6140^{+690}_{-600}$ (or $\mathrm{D/H}=16^{+2}_{-2}\cdot10^{-5}$) that is in tension with the expected protosolar composition ($\mathrm{D/H}_\mathrm{protosolar}=1.67^{+0.25}_{-0.25}\cdot10^{-5}$; \citealt{Asplund_ea_2021}).

The $\mathrm{^{12}CO/^{13}CO}$ isotopologue ratio is constrained at $89^{+2}_{-2}$, in reasonable agreement with the solar-wind value reported by \citet{Lyons_ea_2018}, and within the $2\sigma$ uncertainties of the local interstellar medium (ISM; \citealt{Milam_ea_2005}). The $\mathrm{C^{16}O/C^{17}O}$ value is also in line with the Sun, but is less reliable given the low detection significance of C$^{17}$O. Despite the otherwise solar composition, the $\mathrm{C^{16}O/C^{18}O}$ and $\mathrm{^{16}O/^{18}O}$ ratios of $\sim$\,$800$ unexpectedly imply a depletion of $^{18}$O with respect to the Sun and ISM. 

\subsubsection{Chemical disequilibrium}
\begin{figure*}[ht!]
    \centering
    \includegraphics[width=16cm]{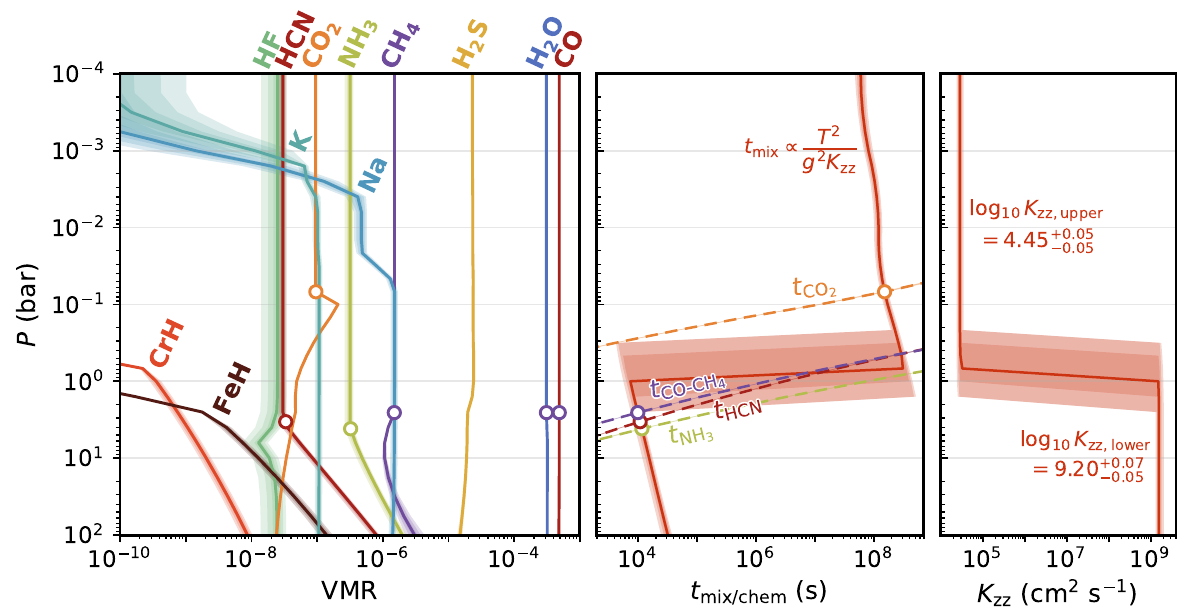}
    \caption{Retrieved vertical profiles of chemistry. \textit{Left panel}: Volume-mixing ratios of molecular and atomic gases. \textit{Middle panel}: Vertical-mixing and chemical timescales. The intersection where mixing becomes more efficient than chemical reactions defines the quench point of the respective gases. \textit{Right panel}: Eddy diffusion coefficient retrieved based on the quenched abundances of CO, CH$_4$, H$_2$O, NH$_3$, and HCN, and separately for CO$_2$.}
    \label{fig:chem_profiles}
\end{figure*}

The spectrum displays strong features of both CH$_4$ and CO, which demonstrates that VHS 1256-1257 b's atmosphere is in a state of chemical disequilibrium. Figure~\ref{fig:chem_profiles} presents the retrieved abundance profiles, the chemical and vertical-mixing timescales, as well as the eddy-diffusion profile. The concentrations of CO, CH$_4$, H$_2$O, NH$_3$, HCN, and CO$_2$ are quenched when mixing becomes more efficient than the corresponding bottleneck reactions \citep{Zahnle_ea_2014}. We note that the quenched CO$_2$ mixing ratio is not directly obtained from its \texttt{FastChem} profile, but derived from the CO, H$_2$O and H$_2$ abundances following \citet{Wogan_ea_2025}. As a result, the constrained CO$_2$ profile in Fig.~\ref{fig:chem_profiles} shows a sudden adjustment at the $\sim$\,$0.1\ \mathrm{bar}$ quench pressure.

The mixing timescale is inversely proportional to the eddy diffusion coefficient, for which the retrieval finds two distinct values of $K_\mathrm{zz,lower}=1.6^{+0.1}_{-0.1}\cdot10^9$ and $K_\mathrm{zz,upper}=2.8^{+0.2}_{-0.2}\cdot10^4\ \mathrm{cm^2\ s^{-1}}$, transitioning between $0.2\lesssim P_{K_\mathrm{zz}}\lesssim2.1\ \mathrm{bar}$, as shown in the middle and right panels of Fig.~\ref{fig:chem_profiles}. These constraints imply that the diffusion coefficient drops by several orders of magnitude when moving up in the atmosphere, which agrees well with the expected transition from a convective to a radiative zone near pressures of $10$--$0.1\ \mathrm{bar}$ \citep{Mukherjee_ea_2022}. The high value of $K_\mathrm{zz,upper}$, which is mainly constrained by the relative abundances of CO and CH$_4$, points towards a vigorous vertical transport of gases in the deep atmosphere. This finding is qualitatively compatible with \citet{Miles_ea_2018}, who also find a high $K_\mathrm{zz}$ of $10^8\ \mathrm{cm^2\ s^{-1}}$ for VHS 1256-1257 b. Moreover, this strong vertical mixing could help to explain the lofty clouds inferred for this atmosphere (Whiteford et al., \textit{subm.}; \citealt{Radcliffe_ea_2026}).

\subsection{Bulk properties} \label{sect:bulk_properties}
\begin{figure}[h!]
    \centering
    \includegraphics[width=\hsize]{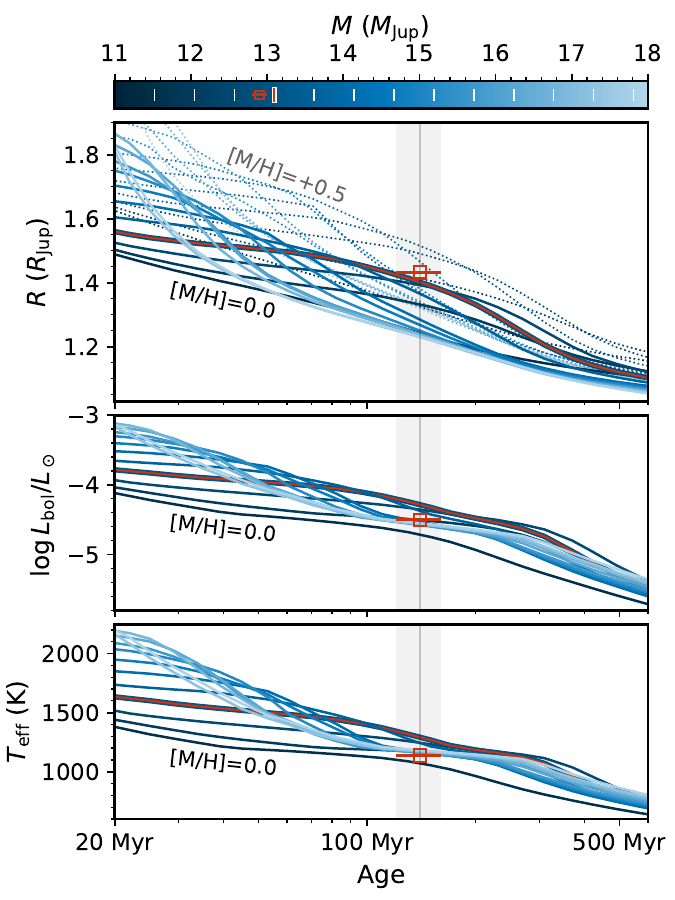}
    \caption{Evolution of the radius, bolometric luminosity, and effective temperature as adopted from the cloudy Sonora Diamondback \textit{hybrid-grav} models \citep{Morley_ea_2024}. Our retrieved $R$, and derived $L_\mathrm{bol}$ and $T_\mathrm{eff}$ are generally consistent with the solar-metallicity evolution tracks at the estimated age of $140\pm20\ \mathrm{Myr}$ \citep{Dupuy_ea_2023}.}
    \label{fig:evolution}
\end{figure}

The retrieval constrains the mass of VHS 1256-1257 b to $M=12.9_{-0.1}^{+0.1}\ M_\mathrm{Jup}$, from a uniform prior between $10$ and $20\ M_\mathrm{Jup}$, and its radius to $R=1.433_{-0.002}^{+0.002}\ R_\mathrm{Jup}$. This works out to a surface gravity of $\log_{10} (g[\mathrm{cm\ s^{-2}}])=4.212_{-0.005}^{+0.005}$, as seen in the corner plot of Fig.~\ref{fig:corner}. We note that our retrieved radius is higher than previous estimates of $\sim$\,$1.3\ R_\mathrm{Jup}$ based on these JWST observations, derived either from the bolometric luminosity \citep{Dupuy_ea_2023} or by fitting the spectral energy distribution (\citealt{Miles_ea_2023,Petrus_ea_2024}; Whiteford et al., \textit{subm.}). This discrepancy can partly be explained by our correction for the $\sim$\,$7$--$10\%$ of flux excluded outside of the extraction aperture, but this would only elevate the previous constraints to $\sim$\,$1.36\ R_\mathrm{Jup}$. As demonstrated in the top panel of Fig.~\ref{fig:evolution}, the Sonora Diamondback cooling track corresponding to the retrieved mass (i.e. $M=13.1\ M_\mathrm{Jup}$; \citealt{Morley_ea_2024}) comes closest to our constrained radius at an age of $140\pm20\ \mathrm{Myr}$ \citep{Dupuy_ea_2023}. Super-solar metallicities can reproduce this radius with a broader range of masses, but we do not find evidence for an enhanced metal content in the atmosphere. We note, however, that both the retrieved radius and mass are strongly correlated with other model parameters and may therefore be biased (see Sect.~\ref{sect:correlation}). 

To further enable a comparison with the evolutionary models and literature, we derived the bolometric luminosity by extending and integrating our two-column model spectra from $0.3$ to $50\ \mathrm{\mu m}$. This results in $\log_{10} (L_\mathrm{bol}/L_\odot)=-4.5044^{+0.0003}_{-0.0003}$, which is somewhat higher than the value of $-4.550^{+0.009}_{-0.009}$ reported by \citet{Miles_ea_2023} as well as the ground-based estimates of $\sim$\,$-4.67$ \citep{Hoch_ea_2022,Petrus_ea_2023}. Our derived $L_\mathrm{bol}$ is nevertheless compatible with the range of luminosities predicted by the self-consistent model libraries considered by \citet{Petrus_ea_2024}. Using the Stefan-Boltzmann law together with the retrieved radius, we derive an effective temperature of $T_\mathrm{eff}=1137^{+1}_{-1}\ \mathrm{K}$, in line with previous studies \citep{Miles_ea_2023,Dupuy_ea_2023,Petrus_ea_2024,Radcliffe_ea_2026}. The lower two panels of Fig.~\ref{fig:evolution} show the overall compatibility between the Sonora Diamondback evolution tracks and our inferred $L_\mathrm{bol}$ and $T_\mathrm{eff}$, even if the small statistical uncertainties do not overlap with the nearest-mass model.

\subsection{Atmospheric structure}
The model comprised of two atmospheric columns provides a significantly better fit compared to a single column. Figure~\ref{fig:1v2_columns} shows that the single-column model struggles to simultaneously reproduce the broader continuum (e.g. $\sim$\,$1.65$, $2.10\ \mathrm{\mu m}$) and the absorption-line depths (e.g. $\sim$\,$3.35$, $4.40$--$5.30\ \mathrm{\mu m}$). The resulting parameter constraints are highly biased, and it is therefore not surprising that evaluation metrics such as the Bayesian Information Criterion show a strong preference for the two-column solution instead ($\Delta\mathrm{BIC}=6200.1$, $\Delta\mathrm{BPICS}=6239.2$; \citealt{Thorngren_ea_2025}). We highlight that the evidence for the two-column solution stems mainly from the wide wavelength range considered in this study. In test retrievals we found that single-column models can still adequately fit the individual gratings, producing spectral residuals at the $1.55$, $0.94$, and $0.71\%$ level (vs $1.58$, $0.96$, $0.68\%$ for the two-column models; see Fig.~\ref{fig:corner_gratings}) for G140H, G235H, and G395H, respectively.  

\begin{figure*}[ht!]
    \centering
    \includegraphics[width=16cm]{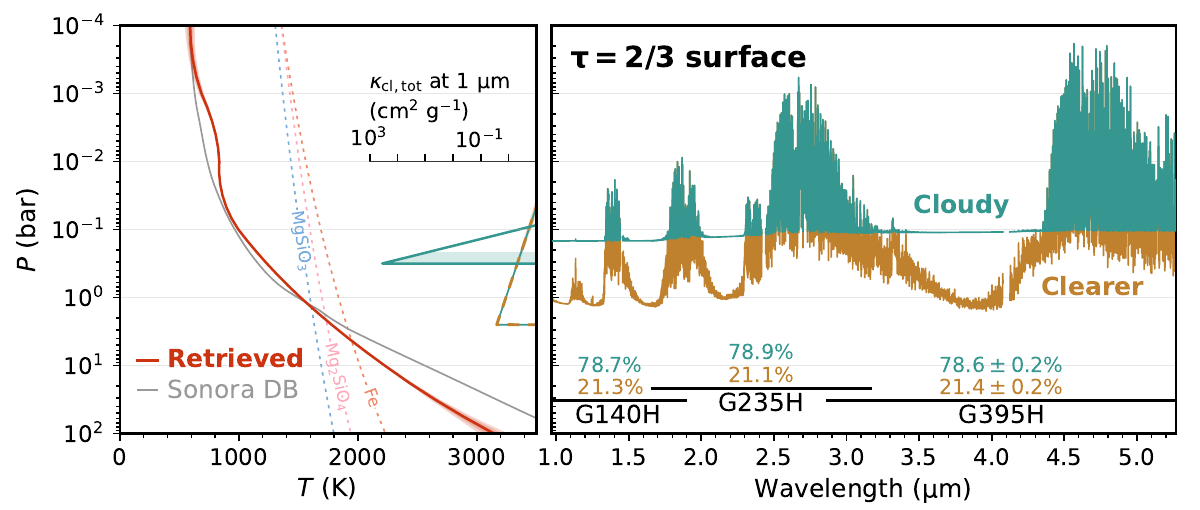}
    \caption{Temperature profile, cloud opacities, and optically-thick surfaces ($\tau=2/3$) in the two-column solution. Certain spectral features seen in Fig.~\ref{fig:bestfit} are primarily formed at the greater depths probed by the clearer column, such as the Na and K lines near $\sim$\,$1.15$ and $1.25\ \mathrm{\mu m}$. Stronger absorption bands of H$_2$O and CO become optically thick at higher altitudes ($P\lesssim$\,$0.1\ \mathrm{bar}$) and therefore originate from both columns, resulting in overlapping $\tau=2/3$ surfaces. The retrieved temperature profile is shown compared to the enstatite, forsterite and iron condensation curves \citep{Visscher_ea_2010}, as well as a representative Sonora Diamondback model \citep{Morley_ea_2024} with $T_\mathrm{eff}=1100\ \mathrm{K}$, $\log_{10}(g[\mathrm{cm\ s^{-2}}])=4.0$, $f_\mathrm{sed}=2$, and $\mathrm{[M/H]}=0$.}
    \label{fig:PT_cloud_profiles}
\end{figure*}

The preferred model consists of a cloudy and clearer atmospheric column that respectively cover $78.6^{+0.2}_{-0.2}$ and $21.4^{+0.2}_{-0.2}\%$ of the visible hemisphere, as noted in Fig.~\ref{fig:PT_cloud_profiles}. We do not recover notable variations in the coverage (see Fig.~\ref{fig:corner}) likely due to the reduced evidence for a two-column solution within individual gratings, as well as the limited rotational phases observed over $1.5\ \mathrm{hr}$ ($\sim$\,$24^\circ$ for $P_\mathrm{rot}=21$--$24\ \mathrm{hr}$; \citealt{Zhou_ea_2020}). Our retrieved coverage differs from the $98.5^{+1.2}_{-2.2}\%$ cloudy coverage Whiteford et al. (\textit{subm.}) constrained for VHS 1256-1257 b using combined NIRSpec and MIRI data, and the $98.745$--$99.225\%$ cover found by \citet{Bowler_ea_2020} through HST variability monitoring. Conversely, our covering fraction is higher than the $62\%$ inferred from the JWST data by \citet{Radcliffe_ea_2026}. However, our result is more consistent with the $\sim$\,$70$--$90\%$ cloud cover found on similar sub-stellar objects \citep{Vos_ea_2023,Molliere_ea_2025,Nasedkin_ea_2025,Zhang_ea_2025}.

In the clearer column, emission can originate from as deep as $\sim$\,$1$--$2\ \mathrm{bar}$, as presented in Fig.~\ref{fig:PT_cloud_profiles}. At those pressures, the opacity from the deep cloud begins to mute the spectral features. The temperature of these photospheric layers reaches a maximum of $\sim$\,$1700\ \mathrm{K}$, thus explaining the flux peak at shorter wavelengths seen in Fig.~\ref{fig:bestfit}. The higher cloud deck in the cloudy column limits the photosphere to $\lesssim0.1\ \mathrm{bar}$, where temperatures are $\lesssim1000\ \mathrm{K}$, resulting in shallower absorption lines and a reddened peak of emission. 

\begin{figure}[h!]
    \centering
    \includegraphics[width=\hsize]{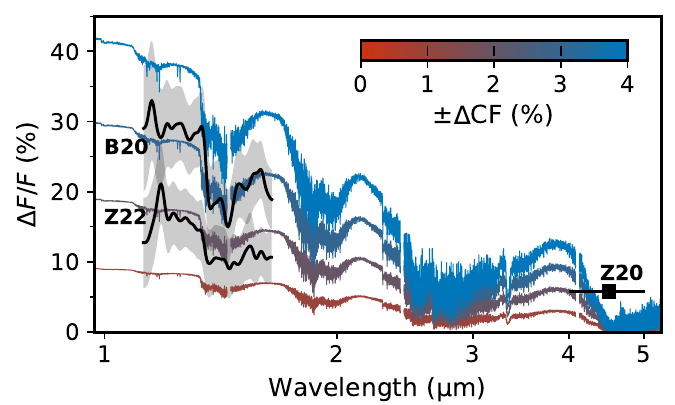}
    \caption{Relative spectral difference resulting from a $\pm\Delta\mathcal{CF}$ percentage change of the inferred cloud-coverage fraction. For reference, we show the HST and Spitzer variability measured for VHS 1256-1257 b in the studies of \citet{Bowler_ea_2020} (as B20), \citet{Zhou_ea_2020} (as Z20), and \citet{Zhou_ea_2022} (as Z22). Small coverage changes of $\pm1$--$3\%$ can generally reproduce the observed variability.}
    \label{fig:coverage_variability}
\end{figure}

Figure~\ref{fig:coverage_variability} illustrates the spectral variability caused by changes to our inferred cloud-coverage fraction. Specifically, it shows the relative difference between a model with a $+\Delta\mathcal{CF}$ increase and one with a $-\Delta\mathcal{CF}$ reduction. Given that the shorter wavelengths are dominated by a smaller clearer patch, perturbations to the covering fraction can greatly enhance or diminish its emission, yielding a high relative flux difference. Small coverage changes of $1$--$3\%$ can therefore reproduce the $10$--$30\%$ level of variability observed with HST/WFC3 for VHS 1256-1257 b \citep{Bowler_ea_2020,Zhou_ea_2022}. The longer-wavelength, Spitzer/IRAC Channel 2 light curve of \citet{Zhou_ea_2020} requires larger coverage modulations, or possible thermo-chemical variations. 

\subsubsection{Thermal and cloud properties}
The left panel of Fig.~\ref{fig:PT_cloud_profiles} shows that the retrieved temperature profile is broadly consistent with a cloudy Sonora Diamondback model at our derived $T_\mathrm{eff}=1100\ \mathrm{K}$ \citep{Morley_ea_2024}. The self-consistent model only predicts a notably higher temperature in the deep atmosphere, where our retrieved profile is primarily constrained by the chemical-equilibrium abundances and not directly by the spectral shape. Compared to the Diamondback profile, we infer a shallower temperature gradient of $\nabla_4=-0.004^{+0.003}_{-0.003}$ at $10^{-2}\ \mathrm{bar}$. The isothermality hinders the absorption of H$_2$O and CO lines near $\sim$\,$2.5$--$2.9$ and $4.5$--$5.0\ \mathrm{\mu m}$, but this is counteracted when the gradient increases at $10^{-3}\ \mathrm{bar}$. Contrary to the inversions detected from CH$_4$ emission on WISE J1935 \citep{Faherty_ea_2024} and SIMP 0136 \citep{Nasedkin_ea_2025}, our upper temperature deviation minimally affects the less opaque CH$_4$ Q-branch at $\sim$\,$3.35\ \mathrm{\mu m}$. Like \citet{Faherty_ea_2024} and \citet{Nasedkin_ea_2025}, we also do not find evidence of H$_3^+$ in the spectral residuals through cross-correlation (nor H$_3$O$^+$ as suggested by \citealt{Helling_ea_2019}). Although this contrasts with the auroral emission on the Solar System giants (e.g. \citealt{Miller_ea_2000}), theoretical work suggests that H$_3^+$ may be efficiently destructed under the sub-stellar atmospheric conditions \citep{Helling_ea_2019,Pineda_ea_2024}. Since the retrieved uncertainty on the thermal gradient is still consistent with $\nabla=0$, we do not view our constraint as evidence for stratospheric, or auroral heating. 

Comparing the retrieved cloud-opacity profiles with the condensation curves of \citet{Visscher_ea_2010}, we find that the lower cloud deck is likely associated with the formation of iron clouds. The upper cloud layer is instead constrained at a higher altitude than the intersections of forsterite (Mg$_2$SiO$_4$) and enstatite (MgSiO$_3$) predict. Similarly, Whiteford et al. (\textit{subm.}) constrain vertically-extended silicate clouds and suggest that the low surface gravity and strong atmospheric mixing of VHS 1256-1257 b work to propel the cloud particles upwards. Then again, our retrieval predicts a rapid opacity decay for the upper cloud and an opacity nearing $\kappa_\mathrm{base,1}\sim$\,$10^3\ \mathrm{cm^2\ g^{-1}}$. Contrary to the expected Rayleigh slope, we also infer a strongly increasing upper-cloud opacity with wavelength. We speculate that these improbable constraints could stem from our simplified cloud- and coverage-parametrisations. A truly heterogeneous surface will be made up of an unlimited number of atmospheric columns, each with their own cloud and thermo-chemical properties \citep{Tan_ea_2025}. If our model was extended with a third patch using an intermediate cloud deck, it may enhance the short-wavelength emission in a similar manner as the positive opacity slope currently does. 

\section{Discussion} \label{sect:discussion}
\subsection{Retrieval robustness and challenges} \label{sect:correlation}
The presented best-fitting model fits for a total of 44 free parameters that do not have independent, unique effects on the generated model spectra. As a result, the retrieval inevitably finds correlations between parameters that can lead to biased constraints. Figure~\ref{fig:correlation} summarises the potential degeneracies of the obtained results using Pearson correlation coefficients. A large number of (anti)-correlations are visible for the mass, radius, coverage fraction, upper cloud deck, and temperature gradients. For example, an increase of the radius can be compensated by a coverage reduction of the clearer column, or a steeper gradient at $0.1\ \mathrm{bar}$ ($\nabla_3$) because that shifts the photosphere to cooler temperatures. We also observe that the quenching of CO$_2$ (i.e. $K_\mathrm{zz,upper}$) is sensitive to the surface gravity (i.e. the mass $M$) due to their influence on the mixing timescale in the upper atmosphere (see Eq.~\ref{eq:t_mix}).

At the chosen constant sampling efficiency of $5\%$ with 1000 live points, the retrieval converges after $\sim$\,$4$ million likelihood evaluations. With our optimisations of the \texttt{pRT} model, this results in a runtime of $\sim$\,$18500$ CPU hours. The necessary usage of a constant efficiency, combined with the limited number of live points for \texttt{MultiNest}, likely yields overconfident posteriors (e.g. \citealt{Buchner_2016,Chubb_ea_2022,Vasist_ea_2023}). To assess the sensitivity of our results to this sampling strategy, we compared our reported constraints to retrievals with 500 live points. The main conclusions are effectively unchanged, still producing an $\sim$\,$80\%$ cloud cover and consistent abundances for carbon, oxygen, sulphur, as well as $^{13}$C and $^{18}$O. The most notable deviations are found for the mass ($10.3^{+0.1}_{-0.1}$ vs $12.9^{+0.1}_{-0.1}\ M_\mathrm{Jup}$) and radius ($1.483^{+0.003}_{-0.003}$ vs $1.433^{+0.002}_{-0.002}\ R_\mathrm{Jup}$), which likely arise from the large number of degeneracies seen in Fig.~\ref{fig:correlation}. Consequently, our reported constraints should be interpreted with this caveat in mind. 

As retrievals move towards a domain of higher spectral resolution, broader wavelength coverage, and complex atmospheric models, other sampling algorithms are likely to improve computational efficiency and accuracy. Our analysis therefore motivates future work that systematically tests samplers such as \texttt{MultiNest} \citep{Feroz_ea_2009}, \texttt{PolyChord} \citep{Handley_ea_2015}, \texttt{dynesty} \citep{Speagle_2020}, \texttt{ultranest} \citep{Buchner_2021}, and \texttt{EMCEE} \citep{Foreman_Mackey_ea_2013}. Such accelerations and validation efforts will be especially important for a combined retrieval analysis of the native-resolution NIRSpec and MIRI spectra of VHS 1256-1257 b, for instance. That endeavour is beyond the scope of this study, but would provide a valuable, comprehensive insight into the (non-uniform) cloud properties and atmospheric chemistry, as well as their interdependence through processes such as oxygen trapping and vertical mixing. 

\subsubsection{Abundance degeneracies} \label{sect:abundance_correlation}
\begin{figure*}[ht!]
    \centering
    \includegraphics[width=15cm]{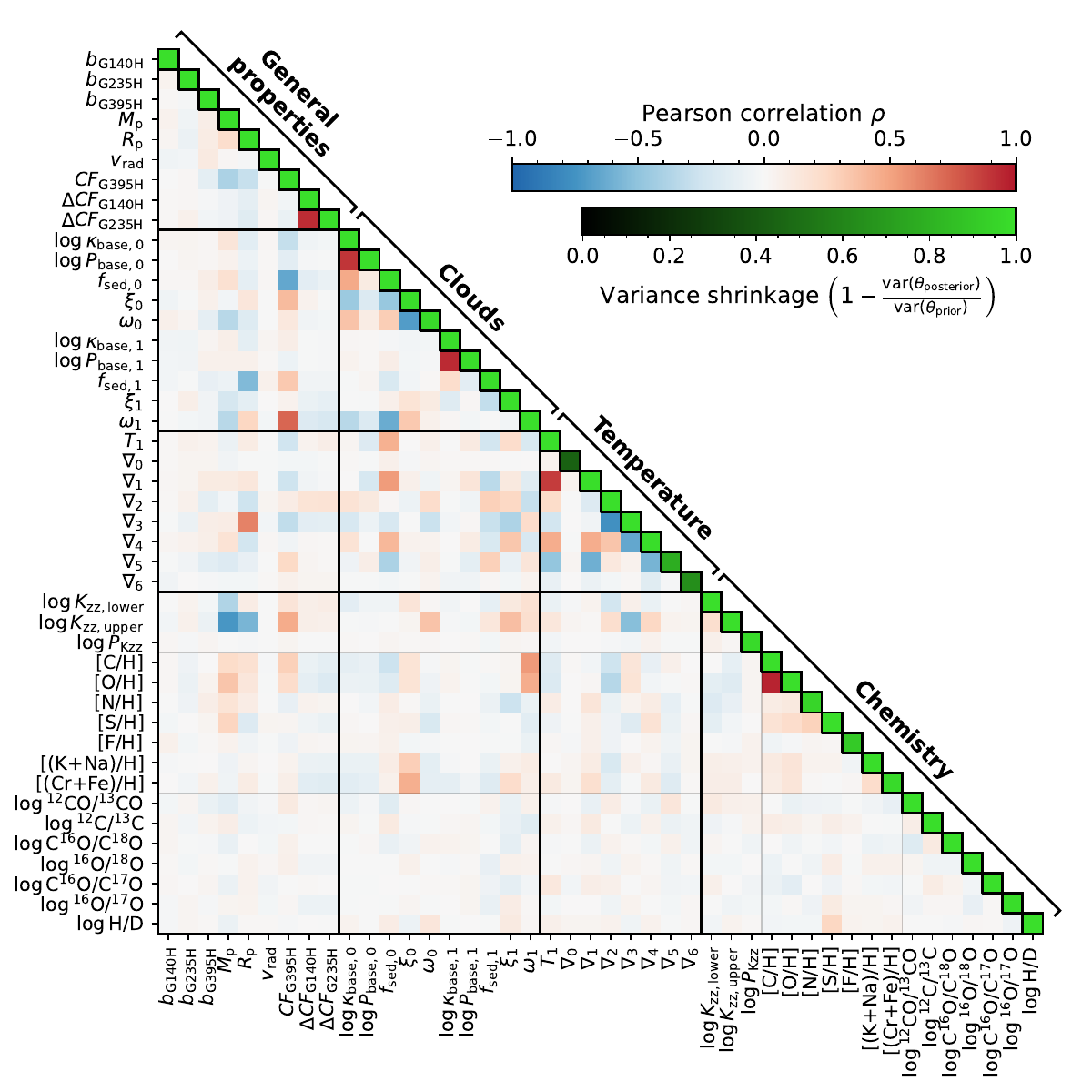}
    \caption{Correlation matrix of the 44 parameters retrieved in the two-column model. The colour of each square shows the Pearson correlation coefficient, $\rho$, of the parameters on the x- and y-axes. Red indicates positive correlation and blue shows anti-correlation. Instead of showing $\rho=1$ on the diagonal, we present the variance shrinkage factors, which measure how well the parameters are constrained relative to their priors. Most parameters are tightly constrained and display a bright green.}
    \label{fig:correlation}
\end{figure*}

Our atmospheric retrieval finds that the concentrations of carbon, oxygen, sulphur, and fluorine in VHS 1256-1257 b are compatible with a solar metallicity. Curiously, the elements nitrogen, potassium, sodium, chromium, and iron are instead constrained at sub-solar abundances. In the correlation matrix of Fig.~\ref{fig:correlation}, we observe that $\mathrm{[(K+Na)/H]}$ and $\mathrm{[(Cr+Fe)/H]}$ are positively correlated with the sedimentation and slope parameters of the bottom cloud deck ($f_\mathrm{sed,0}$, $\xi_0$). Any reduction of the cloud opacity within the clearer column -- by enhancing $f_\mathrm{sed,0}$ or $\xi_0$ -- can be relieved by raising the alkali and metal-hydride concentrations, since they mainly absorb in the bluer G140H grating. Figure~\ref{fig:PT_cloud_profiles} and Table~\ref{tab:parameters} demonstrate that the bottom cloud opacity decays slowly above the condensation base. As such, the $\mathrm{[(K+Na)/H]}$, $\mathrm{[(Cr+Fe)/H]}$, $f_\mathrm{sed,0}$ and $\xi_0$ are possibly reduced in unison to a marginally higher-likelihood solution. 

Figure~\ref{fig:correlation} shows that the total nitrogen abundance, $\mathrm{[N/H]}$, correlates with the atmospheric temperature ($T_1$, $\nabla_1$) and anti-correlates with the eddy diffusion coefficient ($K_\mathrm{zz}$), unlike the carbon and oxygen abundances. The primary driver of the $\mathrm{[N/H]}$ constraint is the volume-mixing ratio of NH$_3$, which favours lower temperatures and higher pressures \citep{Lodders_ea_2002,Marley_ea_2015}. Intrinsic pressure-temperature degeneracies can be broken by measuring multiple molecules with differing abundance dependencies. This is the case for CO, CH$_4$, H$_2$O, and CO$_2$, but not for NH$_3$, because the dominant nitrogen carrier, N$_2$, is not detectable. Since NH$_3$ captures $\lesssim1\%$ of the total nitrogen budget at $P<10\ \mathrm{bar}$ (Fig. 7 of \citealt{de_Regt_ea_2026}), any small underestimation of its abundance can translate into a significantly lower inferred $\mathrm{[N/H]}$.

It is interesting to note the relative absence of degeneracies for the isotopic abundance ratios in Figs.~\ref{fig:correlation} and \ref{fig:corner}, which demonstrates that they are generally unaffected by the other parameter constraints. The $\mathrm{H/D}$ ratio shows only a weak correlation with the sulphur abundance, $\mathrm{[S/H]}$, since HDO and H$_2$S have overlapping opacities near $\sim$\,$3.7\ \mathrm{\mu m}$ \citep{Molliere_ea_2019a}. Analogous to the isotope ratios, the absolute abundances of carbon and oxygen are tied closely together ($\rho=0.9564\pm0.0006$) in order to retain the same atmospheric $\mathrm{C/O}$ ratio. Consequently, elemental and isotopic abundance ratio are less susceptible to degeneracies than the total $[\mathrm{X/H}]$ abundances. Of course, this does not mean that abundance ratios are unbiased because they are still conditioned on a specific data reduction using a parametrised model. 

We constrain a solar $\mathrm{C/O}=0.567^{+0.001}_{-0.001}$ in this work, which is offset from previous JWST retrieval studies of VHS 1256-1257 b. \citet{Gandhi_ea_2023} infer $\mathrm{C/O}=0.69^{+0.01}_{-0.01}$ from the G395H-NRS2 grating-detector pair, and Whiteford et al. (\textit{subm.}) find a bulk $\mathrm{C/O}=0.63^{+0.01}_{-0.02}$ from the combined, downsampled NIRSpec and MIRI spectra. We note that the $\mathrm{C/O}$ ratio reported by \citet{Gandhi_ea_2023} is derived from freely retrieved gaseous abundances, and therefore does not account for the $\sim$\,$10$--$20\%$ of oxygen trapped in silicate-oxide clouds \citep{Calamari_ea_2024}. Whiteford et al. (\textit{subm.}) have corrected for the sequestered oxygen. The previous analyses both utilised the spectral extractions of the initial ERS publication \citep{Miles_ea_2023}, whereas we performed a new data reduction to alleviate spurious continuum oscillations. As seen in Fig.~\ref{fig:ERS_comparison}, the significant differences between the two reductions could reasonably explain discrepant $\mathrm{C/O}$ inferences. Likewise, it seems that the super-solar $^{13}$C, $^{18}$O, and $^{17}$O abundances reported by \citet{Gandhi_ea_2023} (see Fig.~\ref{fig:composition}; $\mathrm{^{12}CO/^{13}CO}=62^{+2}_{-2}$, $\mathrm{C^{16}O/C^{18}O}=425^{+33}_{-28}$, $\mathrm{C^{16}O/C^{17}O}=1010^{+120}_{-110}$) are biased results of the issues in the initial NIRSpec reduction. By applying a cross-correlation analysis to an updated reduction of the MIRI spectra, \citet{Malin_ea_2026} find a CO isotopologue ratio of $\mathrm{^{12}CO/^{13}CO}=78^{+13}_{-10}$ for VHS 1256-1257 b. This is consistent with our constraint of $\mathrm{^{12}CO/^{13}CO}=89^{+2}_{-2}$ from the updated NIRSpec data reduction.

\subsection{Mass of VHS 1256-1257 b}
Previous studies have shown that the luminosity of VHS 1256-1257 b is consistent with both inert and deuterium-fusing masses of $12.0\pm0.1$ and $16\pm1\ M_\mathrm{Jup}$, respectively \citep{Dupuy_ea_2023,Miles_ea_2023}. Using JWST/NIRSpec G395M spectra of the coldest known brown dwarf WISE 0855, \citet{Rowland_ea_2024} reported a detection of CH$_3$D at a protosolar $\mathrm{D/H}$ abundance, indicating a sufficiently low mass to prevent deuterium-fusion. Motivated by this result, we included the isotopologues of HDO and CH$_3$D in our retrieval to investigate whether VHS 1256-1257 b has depleted its primordial deuterium reservoir. 

Although CH$_3$D is the more detectable isotopologue in colder atmospheres \citep{Morley_ea_2019}, \citet{Molliere_ea_2019a} note that the HDO signal increases significantly when quenched CH$_4$ abundances no longer obscure HDO absorption lines near $\sim$\,$3.7\ \mathrm{\mu m}$. In line with this expectation, the auto-correlation functions shown in Fig.~\ref{fig:CCF} (dotted black lines) suggest that HDO should be easier to detect than CH$_3$D in this atmosphere. Despite this, the cross-correlation with the observed spectra does not reveal significant detections of either HDO or CH$_3$D. 

Rather than yielding an upper limit, the retrieval returns a deuterium abundance of $\mathrm{D/H}=(16^{+2}_{-2})\cdot10^{-5}$, an order of magnitude higher than the protosolar value \citep{Asplund_ea_2021}. Since deuterated isotopologues are not convincingly detected, this retrieved abundance should not be taken as evidence of a planetary mass for VHS 1256-1257 b. While the retrieval constrains $\sim$\,$12.9^{+0.1}_{-0.1}\ M_\mathrm{Jup}$, we also caution against interpreting this as evidence for a deuterium-burning mass above the $12.05\ M_\mathrm{Jup}$ limit of \citet{Morley_ea_2024}, in light of the parameter degeneracies found in Fig.~\ref{fig:correlation}.

\subsection{Formation history}
From our atmospheric retrieval, we infer a solar-like carbon isotope ratio of $\mathrm{^{12}C/^{13}C}=89^{+2}_{-2}$ and a tentative oxygen-17 ratio of $\mathrm{^{16}O/^{17}O}=3010^{+320}_{-280}$, whereas $^{18}$O appears depleted relative to the Sun at $\mathrm{^{16}O/^{18}O}\sim$\,$800$. Our derived $\mathrm{^{18}O/^{17}O}=3.9^{+0.5}_{-0.4}$ is consistent with the local ISM value ($4.16\pm0.09$; \citealt{Wouterloot_ea_2008}), but lower than the solar ratio ($5.36\pm0.35$; \citealt{Lyons_ea_2018}). These findings are somewhat surprising as models of galactic chemical evolution predict that the ISM becomes enriched with the heavier isotopes over time \citep{Romano_2022}, a trend that is seemingly confirmed with the compositions of nearby M-dwarf stars \citep{Gonzalez_Picos_ea_2025c}. Since VHS 1256-1257 formed $140\pm20\ \mathrm{Myr}$ ago \citep{Dupuy_ea_2023}, one might expect that the system inherited ISM-like isotope ratios. However, this simple intuition overlooks the role of the formation environment, both within a chemically heterogeneous Milky Way (e.g. \citealt{Wilson_ea_1994,Milam_ea_2005}), and in the interior of star-forming clouds where various fractionation processes alter the accessible isotopic reservoir (see \citealt{Nomura_ea_2023} for a review). For example, the rarer C$^{18}$O isotopologue is less capable of self-shielding from photo-dissociating UV radiation, whereas $^{12}$CO and $^{13}$CO gas become shielded at shallower depths (e.g. \citealt{Visser_ea_2009,Miotello_ea_2014}). This isotope-selective photo-dissociation can strongly elevate the gaseous $\mathrm{C^{16}O/C^{18}O}$ ratios within the star-forming cloud \citep{Nomura_ea_2023}, akin to the super-solar oxygen ratios we infer for VHS 1256-1257 b. The liberated oxygen isotopes are thought to freeze out into water ice (e.g. \citealt{Lyons_ea_2005,Visser_ea_2009}). The accretion of these ices would need to be suppressed to produce an $^{18}$O-depleted composition, but it is puzzling how this can be accomplished for VHS 1256-1257 b.

Although the most likely scenario is a common, top-down formation \citep{Poon_ea_2024}, it remains unclear whether VHS 1256-1257 b shares its composition with the M-dwarf hosts, owing to the lack of abundance measurements for the inner binary. The system is also not part of any young association \citep{Gagne_ea_2026}, thus providing no known coeval objects for comparison. Constraining the composition of the M-dwarf host stars is critical to placing VHS 1256-1257 b's isotope abundances within the system's chemical context and, as a consequence, to better understand its formation history. 

\section{Conclusions} \label{sect:conclusions}
We analysed JWST/NIRSpec spectra of the planetary-mass companion VHS 1256-1257 b to determine the chemical composition of its atmosphere. For this purpose, we performed a new reduction of the ERS integral-field spectra first published by \citet{Miles_ea_2023}, and applied a retrieval framework to infer the atmospheric properties. To our knowledge, this is the first retrieval study of a planetary-mass companion by modeling the combined G140H, G235H, and G395H data at their native spectral resolutions of $\sim$\,$2000$--$4100$. Despite the computational challenge, the model spectrum fits the observations remarkably well across the wide $0.97$--$5.27\ \mathrm{\mu m}$ wavelength range. 

Our retrieval strongly favours ($\Delta\mathrm{BIC}=6200.1$) a two-column solution comprised of cloudy and clearer atmospheric regions that respectively cover $\sim$\,$79$ and $21\%$ of the visible hemisphere. We find that $\pm1$--$3\%$ changes to these coverages can reasonably explain the high $10$--$30\%$ variability seen for VHS 1256-1257 b. This variability mechanism may be examined in more detail through spectroscopic time series, as conducted in JWST GO Program \#3375 (PI: Whiteford). We detect a broad range of gaseous species, including H$_2$O, CO, CH$_4$, CO$_2$, H$_2$S, FeH, HF, NH$_3$, K, Na, $^{13}$CO, C$^{18}$O, H$_2^{18}$O, and find tentative evidence for HCN, CrH, and C$^{17}$O. The deuterated isotopologues HDO and CH$_3$D are not clearly detected, and their retrieved abundances can therefore not distinguish between a planetary or deuterium-fusing mass for VHS 1256-1257 b. The relative abundances of CO, CH$_4$, and H$_2$O suggest efficient vertical mixing of gases at pressures beyond $2\ \mathrm{bar}$ ($K_\mathrm{zz,lower}=1.6^{+0.1}_{-0.1}\cdot10^9\ \mathrm{cm^2\ s^{-1}}$), and a possible reduced efficiency below $0.2\ \mathrm{bar}$, based on a separate quench altitude for CO$_2$ ($K_\mathrm{zz,upper}=2.8^{+0.2}_{-0.2}\cdot10^4\ \mathrm{cm^2\ s^{-1}}$). The orders-of-magnitude change in diffusion coefficients is qualitatively consistent with the expected transition from a convective to a radiative zone \citep{Mukherjee_ea_2022}. 

While the elemental abundances of carbon, oxygen, sulphur, and fluorine indicate a solar metallicity, we attribute the inferred depletions of nitrogen, potassium, sodium, chromium, and iron to parameter degeneracies that are aided by the lack of robust, multi-wavelength detections of these elements. The derived $\mathrm{C/O}=0.567^{+0.001}_{-0.001}$ and $\mathrm{C/S}=20.1^{+0.3}_{-0.3}$ are in good agreement with the solar ratios, and so are the retrieved carbon-13 and oxygen-17 isotope ratios of $\mathrm{^{12}C/^{13}C}=89^{+2}_{-2}$ and $\mathrm{^{16}O/^{17}O}\sim3000$. Since the $\mathrm{^{16}O/^{18}O}$ ratio is inferred at $\sim$\,$800$, VHS 1256-1257 b appears surprisingly depleted in oxygen-18 relative to the Sun. It remains challenging to identify a formation pathway that could have resulted in this elemental and isotopic composition. The misaligned orbits and spin-axis within the VHS 1256-1257 system strongly imply a joint, star-like formation for the inner binary and planetary-mass companion \citep{Poon_ea_2024}. In the absence of known abundances for the M-dwarf hosts, however, we cannot establish a common chemical origin.

The chemical and isotopic composition determined in this work differs substantially from previous JWST retrieval analyses of VHS 1256-1257 b (\citealt{Gandhi_ea_2023}; Whiteford et al., \textit{subm.}). Thus, our results highlight the importance of improved data reduction routines as well as the value of panchromatic, native-resolution retrievals to reveal simultaneous insights of clouds, temperature, and chemistry. Beyond this first ERS spectrum, JWST is powering an increasing number of high-fidelity observations and atmospheric retrievals for directly-imaged companions (e.g. \citealt{Ruffio_ea_2024,Hoch_ea_2025,Zhang_ea_2025,Ruffio_ea_2026,Xuan_ea_2026}) and isolated planetary-mass objects (e.g. \citealt{Faherty_ea_2024,Kuhnle_ea_2025,Nasedkin_ea_2025,Molliere_ea_2025}). Extending such analyses across the full near- and mid-infrared wavelength range will provide further insight into the complex and dynamical atmospheres of sub-stellar objects. 

\begin{acknowledgements}
We thank the anonymous referee for their constructive feedback.
We thank Brendan Bowler and Yifan Zhou for sharing the HST spectral variability measurements of VHS 1256-1257 b. 
S.d.R. and I.S. acknowledge NWO grant OCENW.M.21.010. 
This work used the Dutch national e-infrastructure with the support of the SURF Cooperative using grant no. EINF-9460 and EINF-15873. 
This research has made use of the Astrophysics Data System, funded by NASA under Cooperative Agreement 80NSSC21M00561.
\newline
\textit{Software}: Astropy \citep{Astropy_Collaboration_ea_2022}, corner \citep{Foreman_Mackey_ea_2016}, Matplotlib \citep{Hunter_2007}, NumPy \citep{Harris_ea_2020}, pandas \citep{Pandas_Team_2022}, SciPy \citep{Virtanen_ea_2020}, tol-colors.
\end{acknowledgements}

\bibliographystyle{aa}
\bibliography{references}

\begin{appendix}
\onecolumn
\section{Retrieved parameters}
\begin{table*}[h!]
    \centering
    \caption{Retrieved parameters, their uncertainties, and the chosen priors.}
    \begin{tabular}{lll|r}
        \hline\hline
        \textbf{Parameter} & \textbf{Description} & \textbf{Prior} & \textbf{Retrieved} \\
        \hline
        $10^{b_\mathrm{G140H}}$ & Error inflation for G140H & $\mathcal{U}(0.0,1.0)$ &  $0.0515^{+0.0009}_{-0.0009}$ \\
        $10^{b_\mathrm{G235H}}$ & Error inflation for G235H & $\mathcal{U}(0.0,1.0)$ &  $0.123^{+0.002}_{-0.002}$ \\
        $10^{b_\mathrm{G395H}}$ & Error inflation for G395H & $\mathcal{U}(0.0,1.0)$ &  $0.160^{+0.003}_{-0.003}$ \\
        $M\ [M_\mathrm{Jup}]$ & Mass & $\mathcal{U}(10.0,20.0)$ &  $12.9^{+0.1}_{-0.1}$ \\
        $R\ [R_\mathrm{Jup}]$ & Radius & $\mathcal{U}(1.0,1.7)$ &  $1.433^{+0.002}_{-0.002}$ \\
        $v_\mathrm{rad}\ [\mathrm{km\ s^{-1}}]$ & Radial velocity & $\mathcal{U}(-10.0,0.0)$ &  $-4.11^{+0.07}_{-0.07}$ \\
        $\mathcal{CF}_\mathrm{G395H}$ & Coverage fraction [1] & $\mathcal{U}(0.0,1.0)$ &  $0.214^{+0.002}_{-0.002}$ \\
        $\Delta\mathcal{CF}_\mathrm{G140H}$ & Coverage deviation for G140H [1] & $\mathcal{N}(0.0,0.1)$ &  $-0.0009^{+0.0005}_{-0.0005}$ \\
        $\Delta\mathcal{CF}_\mathrm{G235H}$ & Coverage deviation for G235H [1] & $\mathcal{N}(0.0,0.1)$ &  $-0.0024^{+0.0004}_{-0.0004}$ \\
        \hline
        $\log_{10}\kappa_\mathrm{base,0}\ [\mathrm{cm^2\ g^{-1}}]$ & Cloud base opacity  & $\mathcal{U}(-6.0,3.0)$ &  $-1.53^{+0.03}_{-0.02}$ \\
        $\log_{10}P_\mathrm{base,0}\ [\mathrm{bar}]$               & Cloud base pressure & $\mathcal{U}(-2.0,1.5)$ &  $0.46^{+0.03}_{-0.02}$ \\
        $f_\mathrm{sed,0}$ & Sedimentation rate       & $\mathcal{U}(0.0,10.0)$ &  $0.83^{+0.02}_{-0.03}$ \\
        $\xi_0$            & Wavelength slope         & $\mathcal{N}(0.0,0.5)$ &  $-0.76^{+0.02}_{-0.02}$ \\
        $\omega_0$         & Single-scattering albedo & $\mathcal{U}(0.0,1.0)$ &  $0.24^{+0.04}_{-0.04}$ \\
        $\log_{10}\kappa_\mathrm{base,1}\ [\mathrm{cm^2\ g^{-1}}]$ & Cloud base opacity [2] & $\mathcal{U}(-6.0,3.0)$ &  $>2.61$ \\
        $\log_{10}P_\mathrm{base,1}\ [\mathrm{bar}]$               & Cloud base pressure [2] & $\mathcal{U}(-2.0,1.5)$ &  $-0.471^{+0.009}_{-0.010}$ \\
        $f_\mathrm{sed,1}$ & Sedimentation rate [2]       & $\mathcal{U}(0.0,10.0)$ &  $>9.64$ \\
        $\xi_1$            & Wavelength slope [2]         & $\mathcal{N}(0.0,0.5)$ &  $1.66^{+0.02}_{-0.01}$ \\
        $\omega_1$         & Single-scattering albedo [2] & $\mathcal{U}(0.0,1.0)$ &  $0.39^{+0.02}_{-0.02}$ \\
        \hline
        $T_1\ [\mathrm{K}]$ & Temperature at $10\ \mathrm{bar}$ & $\mathcal{U}(1800,2500)$ &  $2226^{+5}_{-5}$ \\
        $\nabla_0$ & Temperature gradient at $10^2\ \mathrm{bar}$ & $\mathcal{N}(0.15,0.01)$ &  $0.148^{+0.007}_{-0.007}$ \\
        $\nabla_1$ & Temperature gradient at $10\ \mathrm{bar}$ & $\mathcal{N}(0.18,0.04)$ &  $0.153^{+0.003}_{-0.003}$ \\
        $\nabla_2$ & Temperature gradient at $1\ \mathrm{bar}$ & $\mathcal{N}(0.21,0.05)$ &  $0.1746^{+0.0009}_{-0.0010}$ \\
        $\nabla_3$ & Temperature gradient at $10^{-1}\ \mathrm{bar}$ & $\mathcal{N}(0.16,0.06)$ &  $0.190^{+0.001}_{-0.001}$ \\
        $\nabla_4$ & Temperature gradient at $10^{-2}\ \mathrm{bar}$ & $\mathcal{N}(0.08,0.03)$ &  $-0.004^{+0.003}_{-0.003}$ \\
        $\nabla_5$ & Temperature gradient at $10^{-3}\ \mathrm{bar}$ & $\mathcal{N}(0.06,0.02)$ &  $0.146^{+0.010}_{-0.009}$ \\
        $\nabla_6$ & Temperature gradient at $10^{-4}\ \mathrm{bar}$ & $\mathcal{N}(0.0,0.02)$ &  $0.02^{+0.01}_{-0.01}$ \\
        \hline
        $\log_{10}K_\mathrm{zz,lower}\ [\mathrm{cm^2\ s^{-1}}]$ & Eddy diffusion coefficient at $P>P_{K_\mathrm{zz}}$     & $\mathcal{U}(3.0,14.0)$ &  $9.20^{+0.03}_{-0.03}$ \\
        $\log_{10}K_\mathrm{zz,upper}\ [\mathrm{cm^2\ s^{-1}}]$ & Eddy diffusion coefficient at $P\leq P_{K_\mathrm{zz}}$ & $\mathcal{U}(3.0,14.0)$ &  $4.45^{+0.03}_{-0.03}$ \\
        $\log_{10}P_{K_\mathrm{zz}}\ [\mathrm{bar}]$ & Transition pressure of $K_\mathrm{zz}$-profile & $\mathcal{U}(-4.0,2.0)$ &  $-0.2^{+0.3}_{-0.3}$ \\
        $\rm [C/H]$ & Carbon abundance & $\mathcal{N}(0.0,0.2)$ &  $-0.007^{+0.003}_{-0.003}$ \\
        $\rm [O/H]$ & Oxygen abundance & $\mathcal{N}(0.0,0.2)$ &  $0.009^{+0.002}_{-0.002}$ \\
        $\rm [N/H]$ & Nitrogen abundance & $\mathcal{N}(0.0,0.2)$ &  $-0.51^{+0.04}_{-0.05}$ \\
        $\rm [S/H]$ & Sulfur abundance & $\mathcal{N}(0.0,0.2)$ &  $0.029^{+0.007}_{-0.007}$ \\
        $\rm [F/H]$ & Fluorine abundance & $\mathcal{N}(0.0,0.2)$ &  $-0.24^{+0.05}_{-0.06}$ \\
        $\rm [(K+Na)/H]$ & Potassium and sodium abundance & $\mathcal{N}(0.0,0.2)$ &  $-0.28^{+0.02}_{-0.02}$ \\
        $\rm [(Cr+Fe)/H]$ & Chromium and iron abundance & $\mathcal{N}(0.0,0.2)$ &  $-0.65^{+0.03}_{-0.03}$ \\
        $\log_{10}\mathrm{^{12}CO/^{13}CO}$ & $^{13}$CO isotopologue ratio & $\mathcal{U}(1.0,4.0)$ &  $1.95^{+0.01}_{-0.01}$ \\
        $\log_{10}\mathrm{^{12}C/^{13}C}$   & Carbon-13 isotope ratio & $\mathcal{U}(1.0,4.0)$ &  $1.62^{+0.07}_{-0.07}$ \\
        $\log_{10}\mathrm{C^{16}O/C^{18}O}$ & C$^{18}$O isotopologue ratio & $\mathcal{U}(2.0,5.0)$ &  $2.89^{+0.03}_{-0.03}$ \\
        $\log_{10}\mathrm{^{16}O/^{18}O}$   & Oxygen-18 isotope ratio & $\mathcal{U}(2.0,5.0)$ &  $2.94^{+0.02}_{-0.02}$ \\
        $\log_{10}\mathrm{C^{16}O/C^{17}O}$ & C$^{17}$O isotopologue ratio & $\mathcal{U}(2.0,5.0)$ &  $3.48^{+0.04}_{-0.04}$ \\
        $\log_{10}\mathrm{^{16}O/^{17}O}$   & Oxygen-17 isotope ratio & $\mathcal{U}(2.0,5.0)$ &  $3.93^{+0.09}_{-0.08}$ \\
        $\log_{10}\mathrm{H/D}$             & Deuterium isotope ratio & $\mathcal{U}(3.0,7.0)$ &  $3.79^{+0.05}_{-0.04}$ \\
        \hline
    \end{tabular}
    \label{tab:parameters}
    \tablefoot{Uniform priors are indicated with $\mathcal{U}(\mathrm{min,max})$ and Gaussian priors with $\mathcal{N}(\mu,\sigma)$, where $\mu$ and $\sigma$ are the mean and standard deviation. The ``[1]'' (or ``[2]'') label indicates that a parameter affects the atmospheric properties of column 1 (or column 2).}
\end{table*}
\begin{figure*}[ht!]
    \centering
    \includegraphics[width=17cm]{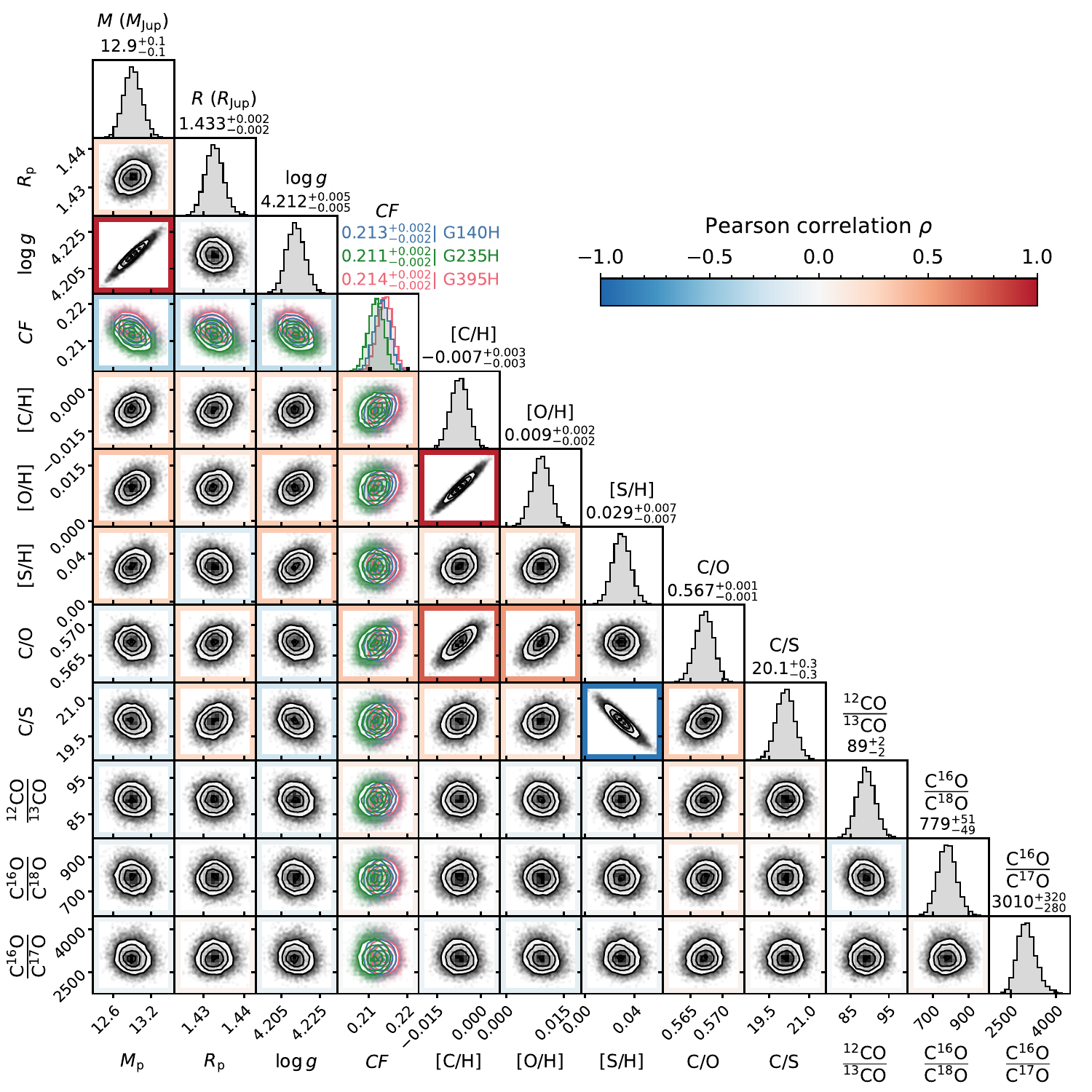}
    \caption{Selected posterior distributions of retrieved, and derived ($\log_{10}(\textit{g})$, $\mathrm{C/O}$, $\mathrm{C/S}$) parameters. The coverage fraction, $\mathcal{CF}$, is plotted for each grating. Similar to Fig.~\ref{fig:correlation}, we show the Pearson correlation coefficients as the border colour of each panel. Importantly, the isotope ratios exhibit minimal correlation with the other parameters, thereby giving confidence in the constrained results.}
    \label{fig:corner}
\end{figure*}

\begin{figure*}[ht!]
    \centering
    \includegraphics[width=17cm]{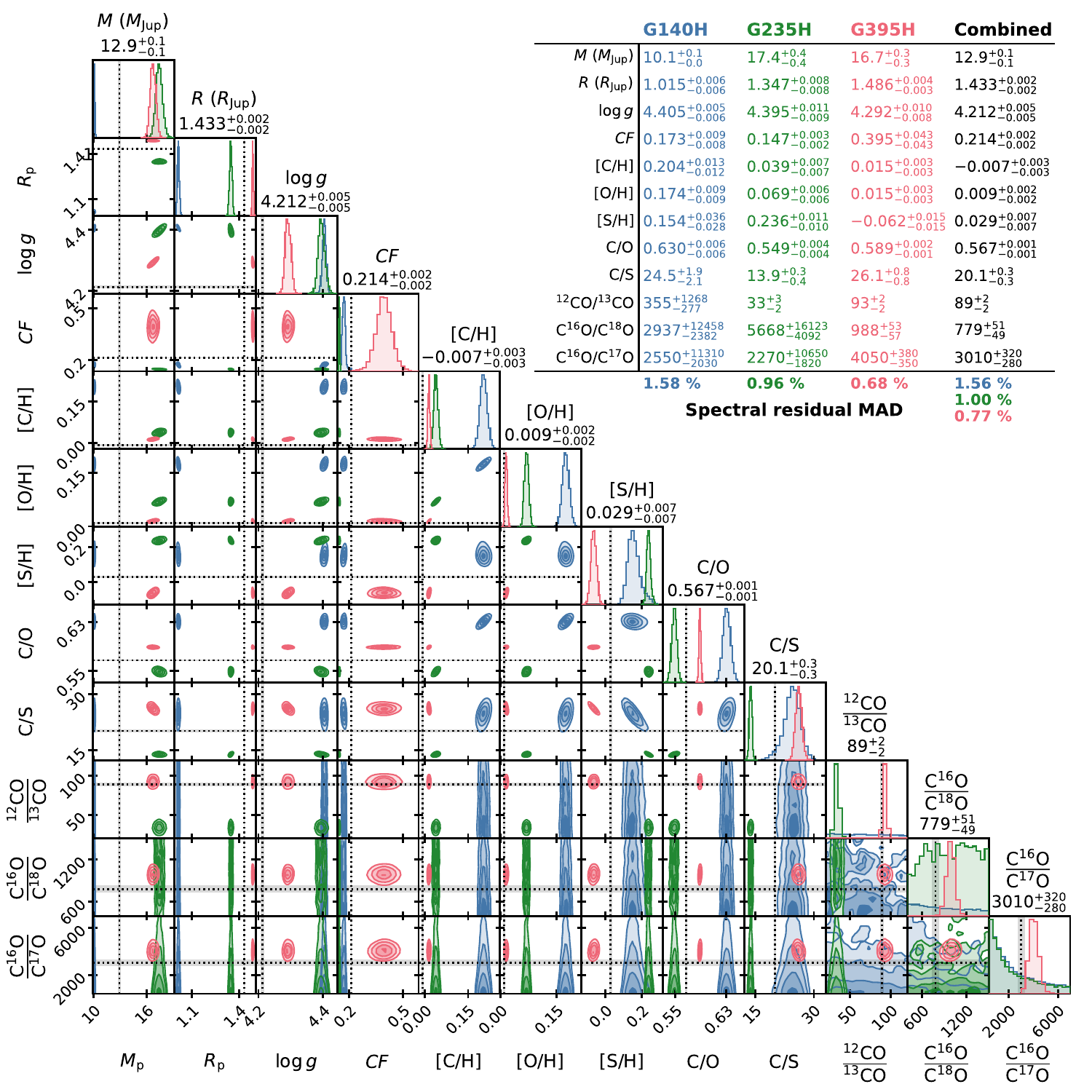}
    \caption{Posterior distributions from separate retrievals of the three gratings: G140H (red), G235H (green), and G395H (blue). For comparison, we show the results from the combined retrieval as a dotted line with $1\sigma$-shading in each panel. A table of retrieved and derived values is also presented in the upper right corner.}
    \label{fig:corner_gratings}
\end{figure*}

\clearpage
\section{Best-fitting spectra}
\begin{figure*}[ht!]
    \centering
    \includegraphics[width=17cm]{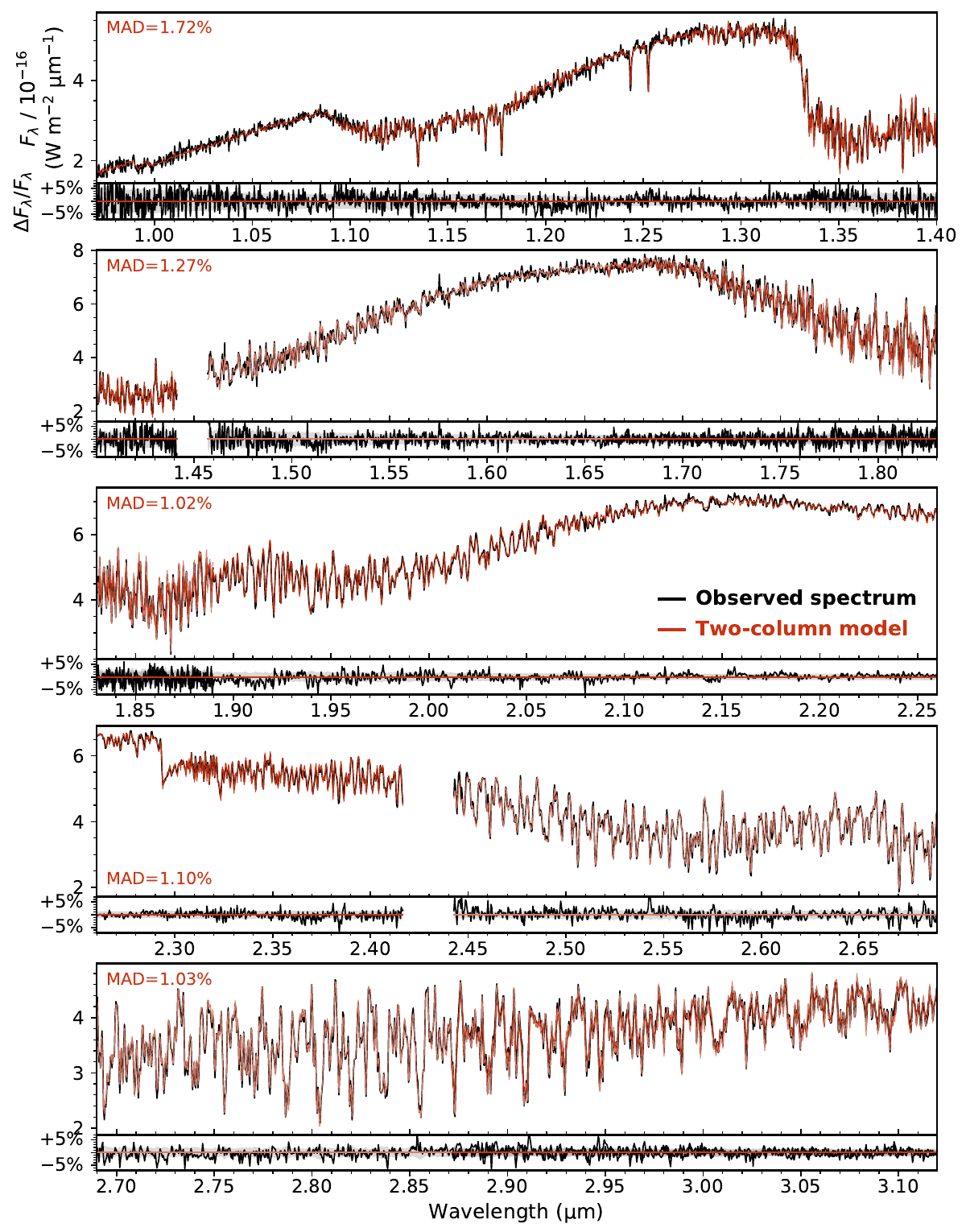}
    \caption{Zoomed-in frames of the observed (black) and best-fitting model spectra (red). The two-column atmospheric model gives a good fit to the data, with residual median absolute deviations (MAD) at the $\sim$\,$1\%$ level. The panels also show the overlap between the spectral gratings ($1.66$--$1.89$, $2.87$--$3.17\ \mathrm{\mu m}$), and we note that their different resolving powers are apparent from the imperfect match of the model spectra. The inflated flux errors are displayed as gray shading behind the residuals.}
    \label{fig:bestfit_zooms}
\end{figure*}
\begin{figure*}[ht!]
    \addtocounter{figure}{-1}
    \centering
    \includegraphics[width=17cm]{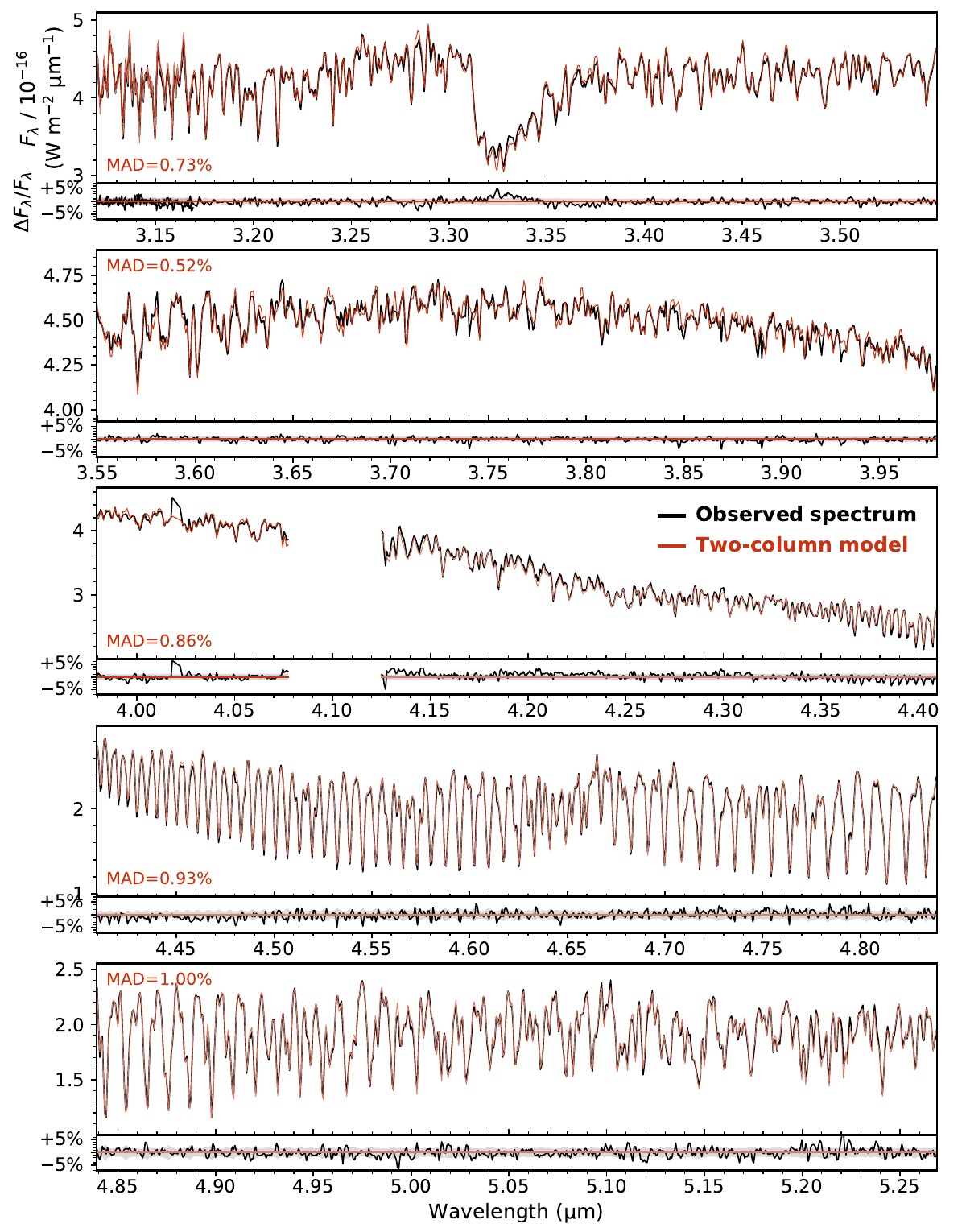}
    \caption{Continued.}
\end{figure*}

\clearpage
\begin{figure*}[ht!]
    \centering
    \includegraphics[width=17cm]{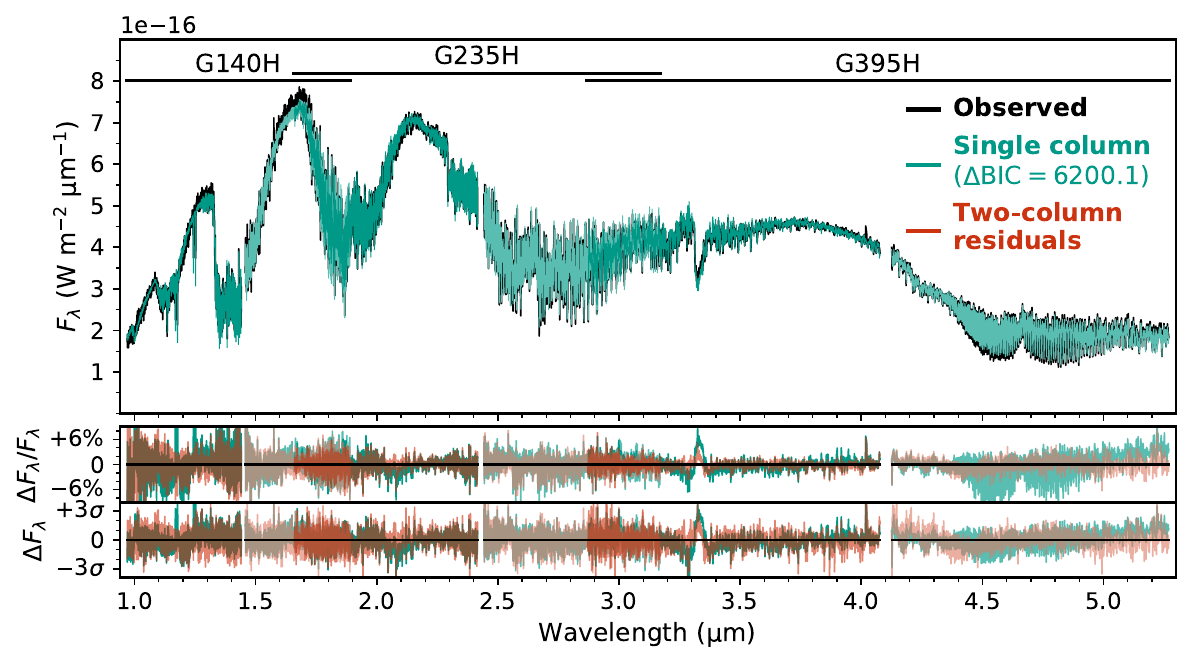}
    \caption{Best-fitting model spectrum (teal) using a single atmospheric column to fit the NIRSpec data (black) of VHS 1256-1257 b. As a reference, we show the residuals from the preferred two-column solution in the lower panel in red. The single-column model is unable to reproduce the continuum and absorption line depths simultaneously.}
    \label{fig:1v2_columns}
\end{figure*}

\section{Utilised line-opacity data}
\begin{table}[ht!]
    \centering
    \caption{References for the utilised line-opacity data.}
    \label{tab:opacity_refs}
    \begin{tabular}{r|l}
        \hline\hline
        H$_2$O                   & \citet{Polyansky_ea_2018} \\
        H$_2^{18}$O, H$_2^{17}$O & \citet{Polyansky_ea_2017} \\
        HDO                      & \citet{Voronin_ea_2010} \\
        CO, $^{13}$CO, C$^{18}$O, C$^{17}$O & \citet{Li_ea_2015} \\
        CH$_4$                   & \citet{Yurchenko_ea_2024} \\
        $^{13}$CH$_4$, CH$_3$D   & \citet{Gordon_ea_2026} \\
        CO$_2$, $^{13}$CO$_2$, $^{16}$OC$^{18}$O & \citet{Yurchenko_ea_2026} \\
        NH$_3$                   & \citet{Coles_ea_2019} \\
        H$_2$S                   & \citet{Azzam_ea_2016,Chubb_ea_2018} \\
        HF                       & \citet{Li_ea_2013,Coxon_ea_2015}; \\
        {}                       & \citet{Somogyi_ea_2021} \\
        HCN                      & \citet{Barber_ea_2014} \\
        CrH, FeH                 & \citet{Bernath_2020} \\
        SiO                      & \citet{Yurchenko_ea_2022} \\
        K, Na                    & \citet{Kurucz_2018} \\
        \hline
    \end{tabular}
\end{table}
\end{appendix}

\end{document}